\documentclass[preprint,nofootinbib,aps,superscriptaddress]{revtex4} 
\usepackage[dvips]{graphicx}
\usepackage{epsfig}
\usepackage{amsmath,amsfonts,amssymb,amsthm}
\usepackage{verbatim}
\usepackage{psfrag}
\usepackage{bm}
\usepackage{bbm}
\usepackage{color}
\usepackage{slashed}

\usepackage{epsf,epsfig}
\usepackage{graphics}
\usepackage{axodraw}

\def\be{\begin{equation}}
\def\ee{\end{equation}}
\def\bea{\begin{eqnarray}}
\def\eea{\end{eqnarray}}

\begin{document}

\hfill TUM-HEP-897-13

\hfill NPAC-13-05


\title{Cuts, Cancellations and the Closed Time Path:\\The Soft
Leptogenesis Example}

\author{Bjorn Garbrecht}
\email{garbrecht@tum.de}
\affiliation{Physik Department T70, James-Franck-Stra{\ss}e,\\
Technische Universit\"at M\"unchen, 85748 Garching, Germany}
\author{Michael J. Ramsey-Musolf}
\email{mjrm@physics.wisc.edu}
\affiliation{University of Wisconsin-Madison, Madison,
WI 53706, USA} 
\affiliation{University of Massachusetts-Amherst, Amherst, MA 01003 USA}
\affiliation{California Institute of
Technology, Pasadena, CA 91125 USA}


\begin{abstract}

By including all leading quantum-statistical effects at finite temperature,
we show that no net asymmetry of leptons and sleptons is generated
from soft leptogenesis,
save the possible contribution from the resonant mixing of sneutrinos. 
This result contrasts with different conclusions appearing in the literature that are based on an
incomplete inclusion of quantum statistics. We discuss vertex and wave-function diagrams as well as all different
possible kinematic cuts that nominally lead to CP-violating asymmetries.
The present example of soft leptogenesis
may therefore serve as a paradigm in order to identify more generally applicable caveats relevant to alternative
scenarios for baryogenesis and leptogenesis, and it may provide
useful guidance in constructing viable models.

\end{abstract}

\thispagestyle{empty}



\vspace{0.4cm}

%
%
%
%
\allowdisplaybreaks[2]

\maketitle

\section{Introduction}

For many extensions of the Standard Model (SM),
additional degrees of freedom and
parameters entail the
possibility of new CP-violating (CPV) phases besides the observed one present in the
CKM matrix. In the absence of further modifications, the SM fails to
explain the observed baryon asymmetry of the Universe (BAU), so it is interesting
to investigate whether the new degrees of freedom and phases may  remedy this
shortcoming. It would be particularly interesting if the
new particles and phases responsible for the BAU
were
experimentally accessible through high energy collisions close to the electroweak scale
(or not too far above it)
or through
the observation of permanent electric dipole moments (EDMs) (for recent reviews and extensive references, see Refs.~\cite{Engel:2013lsa,Morrissey:2012db}). 

A particularly rich model with such phenomenological and cosmological prospects
is the Minimal Supersymmetric Standard Model supplemented by right-handed
singlet neutrinos ($\nu$MSSM). New CPV phases
can be present within the triscalar couplings and the masses that lead to soft supersymmetry
breaking in conjunction with the masses and couplings in the superpotential. This model predicts supersymmetric particles  with masses close to electroweak scale, new CPV signals, and
possibly  an explanation for the emergence of the BAU through the mechanism
of soft leptogenesis~\cite{Grossman:2003jv,D'Ambrosio:2003wy,Grossman:2004dz,Fong:2009iu,Fong:2010qh,Fong:2011yx}.

Of course, the MSSM
is of great interest because it offers a solution to the hierarchy
problem, provides for radiative electroweak symmetry-breaking, and contains particle candidates for Dark Matter.  Moreover,
many of its features are paradigmatic for other extensions of the SM.
Therefore, it is a highly suitable arena for developing theoretical techniques needed for robust computations 
of the BAU, identifying related observable CPV effects, and delineating benchmarks for future experimental CPV searches.

These remarks outline the context of the present paper.
Using the example of soft leptogenesis in the $\nu$MSSM, we study the question whether experimentally
accessible CPV may be linked to non-resonant variants
of baryogenesis or leptogenesis from out-of-equilibrium decays or inverse
decays close to the electroweak scale. Successful baryogenesis from out-of-equilibrium decays,
its occurrence at low temperatures (close to the electroweak
scale) and experimentally observable CPV are often incompatible requirements.
However, it has been proposed that
even in absence of resonant enhancement, soft leptogenesis
is a viable mechanism for baryogenesis at relatively low temperatures, with 
the mass of the decaying singlet neutrino and the temperature at which leptogenesis takes place
possibly being of order of the electroweak scale~\cite{Fong:2009iu,Fong:2010qh,Fong:2011yx}.

At first glance, this may not appear possible, as 
the diagrams for vacuum decays of the singlet neutrinos
that lead to the lepton and slepton asymmetries do not involve (s)lepton
number violation in the internal lines. It follows from the CPT theorem that in the 
vacuum, the produced asymmetries in leptons and sleptons are precisely opposite~\cite{Fong:2009iu}.
Assuming a fast equilibration between particles and sparticles, no net lepton
number
is produced. It has been argued in Refs.~\cite{Fong:2009iu,Fong:2010qh,Fong:2011yx} that a loophole
opens at finite temperature, wherein phase space modifications associated with 
Fermi suppression (leptons)  and Bose enhancement (sleptons) render the vacuum cancellation ineffective. 
Should this loophole, indeed, prove to be viable, one could anticipate a variety 
more phenomenologically interesting models of baryogenesis
from out-of-equilibrium decays besides soft leptogenesis that link the BAU
with experimentally accessible new particles and  CPV signals.

Assessing the viability of the proposal requires careful scrutiny of the unitary
evolution of the full set of quantum statistical states as well as thermal effects.
The quantitative analysis in Ref.~\cite{Fong:2009iu} only
takes account of the quantum statistical corrections for the external states but not those appearing in the
loops. However, when calculating the imaginary parts of the loop diagrams using Cutkosky 
rules, it is immediately obvious (and also well-known) that the imaginary part 
relevant for the asymmetry arises from momentum regions where particles on internal lines
are on-shell. It is, thus, natural to ask whether one must also  consider
quantum statistical enhancement and suppression factors for the internal propagators
as well. In what follows, we demonstrate that it is, indeed, necessary to consider the thermal statistical 
factors for internal lines and that doing so closes the loophole proposed in Refs.~\cite{Fong:2009iu,Fong:2010qh,Fong:2011yx} .

 Although we focus on the model of soft leptogenesis for concreteness, we emphasize that our analysis generalizes to other scenarios,  illustrating important features that should be taken into account whenever 
an asymmetry is supposed to be generated from out-of-equilibrium dynamics in a spatially homogeneous background. For a recent example of the latter, see Ref.~\cite{Hall:2010jx}, wherein a crucial cancellation of the final asymmetry is missed. In this respect, the present work provides a useful guideline
for building successful models of baryogenesis and leptogenesis.

To set the stage for our discussion, we note that in the conventional approach to leptogenesis followed in Refs.~\cite{Fong:2009iu,Fong:2010qh,Fong:2011yx} , which combines $S$-matrix elements from
quantum theory with Boltzmann equations from classical physics, no explicit set of
rules for correctly including  these quantum
statistical effects has been worked out. In particular,
it is not immediately clear whether the on-shell contributions in internal propagators
of loop diagrams
are already accounted for through subsequent tree-level scatterings that are described
in the Boltzmann equations. This issue is of pivotal relevance for the correct 
calculation of the asymmetry. In standard leptogenesis with classical statistics
(which is a good approximation in the strong washout regime), one may take account
of this through the procedure of real intermediate state (RIS) subtraction.

More specifically, when an external particle attaches
to a loop of two or more particles, potentially
CPV cuts occur from the momentum region where the loop
particles are on-shell.\footnote{At zero temperature, this implies that
the sum of the masses of the loop particles must be below the mass of the external particle. At finite temperature, also kinematically allowed
crossings of the internal propagators contribute to the cuts.}  In order to ensure a unitary evolution,
one must also include in the Boltzmann equations scattering diagrams where the unstable particle appears
in an internal line. To avoid double-counting, the RIS must then be subtracted from the
scattering rates in such a way that no charge asymmetry is generated
in equilibrium.  Given the
care required in identifying and performing the full set of RIS subtractions, it is perhaps not surprising that its implementations in Refs.~\cite{Fong:2009iu,Hall:2010jx} were not complete and, as a result, yield spurious, non-vanishing  asymmetries.

In performing the
RIS subtraction, one encounters two cases, corresponding to whether or not the on-shell intermediate particle is in equilibrium.
The present example of soft 
leptogenesis illustrates both situations. Looking ahead to our detailed calculation in Sections \ref{sec:Vertex} and \ref{section:wv}, we summarize the key physics for each. 
\begin{itemize}
\item[(1)] For the vertex contributions,
the intermediate on-shell particle is an out-of equilibrium singlet neutrino $N$ (see  Fig.~\ref{fig:vertex:cuts}). In this case,
even when the subtraction of RIS is performed correctly\footnote{The subtraction is effected either by systematic
derivation as performed here or simply by imposing vanishing net CPV
rates in equilibrium.}, there occurs a cancellation due to opposite
asymmetries in leptons $\ell$ and sleptons $\widetilde \ell$.
A derivation of this cancellation requires the correct inclusion
of all quantum statistical factors for the external
states as well as for intermediate on-shell particles
(see again Fig.~\ref{fig:vertex:cuts}). Quantum statistics has
only partly been accounted for in Ref.~\cite{Fong:2009iu}, which is why
the precise cancellation is missed there.
\item[(2)] On the other hand, when the RIS that is to be subtracted corresponds to a 
particle that is in equilibrium, such as the scalar Higgs doublet $H_1$
in Section~\ref{section:wv} (see Fig.~\ref{fig:offhiggswv}),
no asymmetry is generated in first place. This type of
subtraction of equilibrium RIS and the consequent vanishing of
the CPV asymmetry has been missed in the context of a
different model in
Ref.~\cite{Hall:2010jx}
\end{itemize}

Alternatively, there exists a way of deriving the leptogenesis kinetic equations
following a set of rules that intrinsically
respect the unitary evolution of the system while sidestepping the pitfalls of the RIS procedure: the Closed Time Path (CTP)
formalism~\cite{Schwinger:1960qe,Keldysh:1964ud,Mahanthappa:1962ex,Bakshi:1962dv,Bakshi:1963bn,Calzetta:1986cq}. Rather than formulating the problem in terms of $S$-matrix elements
and classical particle distribution functions, the evolution of Green functions of
the quantum fields is calculated. In particular, the imaginary
parts of self-energies correspond to the inclusive decay and production rates that
are necessary in order to track the evolution of the asymmetry. This
way, the somewhat heuristic procedure of RIS subtraction can be 
avoided~\cite{Buchmuller:2000nd,De Simone:2007rw,Garny:2009rv,Garny:2009qn,Anisimov:2010aq,Garny:2010nj,Beneke:2010wd,Beneke:2010dz,Garny:2010nz,Garbrecht:2010sz,Anisimov:2010dk,Garbrecht:2011aw,Garny:2011hg,Garbrecht:2012qv,Garbrecht:2012pq,Frossard:2012pc}.

In the present work, we calculate the source terms for the asymmetry using the CTP 
formalism. As our main result, we demonstrate that the resulting asymmetry of the lepton
number vanishes even when taking into account  quantum statistical corrections. In particular, the corrections
associated with the internal lines precisely cancel those associated with the final states that are included in Refs.~\cite{Fong:2009iu,Fong:2010qh,Fong:2011yx}. Consequently, the sum of the lepton and slepton asymmetries is zero\footnote{One should mention however, that at temperatures above
$10^7\,{\rm GeV}$, the equilibration of leptons and sleptons  becomes
ineffective, because it is suppressed by a helicity flip of
the mediating gaugino~\cite{Fong:2010bv}. As leptons and sleptons suffer
different washout rates, a net asymmetry can emerge in such a situation.}.
We also note that
while we perform the explicit calculations in the context of soft leptogenesis, our
findings can be straightforwardly generalized to other
conceivable variants of baryogenesis or leptogenesis that
exhibit similar CPV diagrammatic cuts.

The plan of this paper is as follows: In Section~\ref{sec:SoftLepto},
we present the Lagrangian that is relevant for soft leptogenesis.
Furthermore, we define the collision terms that enter kinetic equations,
which can be used to calculate the lepton asymmetry. The
main subject of the present paper are the CPV contributions
to the collision terms and  the crucial cancellations that these
exhibit. In Section~\ref{sec:Vertex}, we start with the general
CTP expression for the vertex-type self-energy that may lead to
the production of asymmetries. We then describe the strategy for extracting
the particular CPV contributions from this self-energy.
In the remaining Subsections of Section~\ref{sec:Vertex}, we consider
various kinematic cuts and demonstrate that in each case,
there is a cancellation of the asymmetries. The corresponding calculations
for the wave-function type self-energy are presented in
Section~\ref{section:wv}. As we discuss in Section~\ref{sec:RIS}, some of our results can be related to
the RIS subtraction procedure that, as noted above, 
is routinely used in standard calculations for leptogenesis . Conclusions are
presented in Section~\ref{sec:conclusions}.

\section{Kinetic Equations in the CTP Framework}
\label{sec:SoftLepto}

The authors of Refs.~\cite{Fong:2009iu,Fong:2010qh,Fong:2011yx} considered the impact of CPV phases that appear in the
soft SUSY-breaking singlet sneutrino (${\widetilde N}$) and wino (${\widetilde W}$) mass terms as well as the trilinear interactions involving the slepton and Higgs doublets and ${\widetilde N}$. For pedagogical purposes, we will consider a different source of CPV associated with the relative phase of the bino ${\widetilde B}$ mass term and the supersymmetric $\mu$ parameter, though the logic and result (vanishing total lepton number asymmetry)  in both cases will be the same. Consequently, in the Lagrangian below, we do not include singlet sneutrinos and winos.
Singlet sneutrinos are of particular interest when their $b$-term is small
compared to their lepton-number violating mass and
they induce a splitting into two almost degenerate  mass
eigenstates~\cite{Grossman:2003jv,D'Ambrosio:2003wy,Grossman:2004dz}.
This opens up
the possibility for a variant of resonant leptogenesis. However, in the
present paper, we restrict our analysis to the non-resonant regime. Since 
singlet sneutrinos can play a role for non-resonant CPV that is analogous
to the one of the singlet neutrinos, the analysis presented here for fermionic 
singlets generalizes in a straightforward manner to the bosonic case. Thus,  we do 
not reiterate it here. Similarly, the effects from bino-mediated interactions in soft leptogenesis that
we present in this paper are analogous to those mediated by winos. For simplicity, we
therefore omit the discussion of the latter.

After suitable field redefinitions through rephasings the $\nu$MSSM, mass and interaction terms relevant to our analysis are
\begin{align}
{\cal L} \supset& -(m_{Hu}^2+\mu^2) H_u^\dagger H_u -(m_{Hd}^2+\mu^2) H_d^\dagger H_d
-b\left(H_u^T H_d + H_u^\dagger H_d^*\right)
\notag\\
-&\mu \bar \Psi_{\widetilde H^+}\Psi_{\widetilde H^+}
-\mu \bar \Psi_{\widetilde H^0}\Psi_{\widetilde H^0}
-\frac 12 M_1 \bar\Psi_{\widetilde B} \Psi_{\widetilde B}
-\frac 12 m_{N}\bar\Psi_N \Psi_N
-{\widetilde \ell}^\dagger m_{\widetilde \ell}^2 \widetilde \ell
\notag\\
-&\frac{g_1}{\sqrt 2}
\left[
\bar \Psi_{\widetilde H^+}\left(
-{H_d^-}^*P_{\rm L}+{\rm e}^{\rm i \phi_\mu} H_u^+ P_{\rm R}
\right)\Psi_{\widetilde B}
+\bar \Psi_{\widetilde H^0}\left(
-{H_d^0}^*P_{\rm L}-{\rm e}^{\rm i \phi_\mu} H_u^0 P_{\rm R}
\right)\Psi_{\widetilde B}
+{\rm h.c.}
\right]
\notag\\
+&\frac{g_1}{\sqrt2}\left[
\widetilde\nu_{\rm L}^*\bar\Psi_{\widetilde B}P_{\rm L}\nu_{\rm L}
+\widetilde e^{-*}_{\rm L}\bar\Psi_{\widetilde B}P_{\rm L}e^-_{\rm L}
+{\rm h.c.}
\right]
\notag\\
-&\left[Y{\rm e}^{-{\rm i}\phi_Y}
\left( H_u^+ \bar \Psi_N  P_{\rm L} e_{\rm L}
-H_u^0 \bar \Psi_N  P_{\rm L} \nu_{\rm L}
\right)
+{\rm h.c.}
\right]
\notag\\
+&\left[Y{\rm e}^{-{\rm i}\phi_\mu-{\rm i}\phi_Y}
\left(
\widetilde \nu_{\rm L}
\bar\nu_{\rm R}P_{\rm L}\Psi_{\widetilde H^0}
+\widetilde e^-_{\rm L}
\bar\nu_{\rm R}P_{\rm L}\Psi_{\widetilde H^+}
\right)
+{\rm h.c.}
\right]
\label{TheLagrangian}
\,,
\end{align}
where
\begin{eqnarray}
\Psi_{\widetilde H^+}=
\left(
\begin{array}{c}
\widetilde H_u^+\\
\widetilde H_d^{-^\dagger}
\end{array}
\right)\,,\quad
\Psi_{\widetilde H^0}=
\left(
\begin{array}{c}
-\widetilde H_u^0\\
\widetilde H_d^{0^\dagger}
\end{array}
\right)\,,\quad
\Psi_{\widetilde B}
=
\left(
\begin{array}{c}
\widetilde B\\
\widetilde B^\dagger
\end{array}
\right)\,,\quad
\Psi_{N}
=
\left(
\begin{array}{c}
N\\
N^\dagger
\end{array}
\right)
\,.
\end{eqnarray}

Within the symmetric electroweak phase,
the scalar Higgs fields are transformed to a
diagonal basis through
\begin{equation}
\left(
\begin{array}{c}
H_u^+\\{H_d^-}^*
\end{array}
\right)
=\left(
\begin{array}{cc}
\cos\alpha & \sin\alpha\\
-\sin\alpha & \cos\alpha
\end{array}
\right)
\left(
\begin{array}{c}
H_1^+\\H_2^+
\end{array}
\right)
\,,
\quad
\left(
\begin{array}{c}
H_u^0\\{H_d^0}^*
\end{array}
\right)
=\left(
\begin{array}{cc}
\cos\alpha & -\sin\alpha\\
\sin\alpha & \cos\alpha
\end{array}
\right)
\left(
\begin{array}{c}
H_1^0\\H_2^0
\end{array}
\right)\,,
\end{equation}
where
\begin{equation}
\tan 2\alpha=\frac{2 b}{m_{Hu}^2-m_{Hd}^2}\,.
\end{equation}
The mass-square eigenvalues for $H_{1,2}^{0,+}$ are
\begin{equation}
m^2_{H{1,2}}
=\frac 12\left(m_{Hu}^2 + m_{Hd}^2 \pm \sqrt{(m_{Hu}^2 - m_{Hd}^2)^2+4b^2}\right)\,.
\end{equation}
For $m_{Hd}^2-m_{Hu}^2 \gg |b|$, as  is often assumed in
particular MSSM scenarios, the mixing angle is approximately given by
$\sin\alpha\approx 2b/(m_{Hu}^2-m_{Hd}^2)$, and the mass squares are
$m_{H1}^2\approx m_{Hu}^2$ and $m_{H2}^2\approx m_{Hd}^2$. 

The kinetic equations for soft leptogenesis need to track the distribution
of singlet neutrinos, $f_N(\mathbf k)$ and  the asymmetries
of leptons $f_\ell(\mathbf k)-\bar f_\ell(\mathbf k)$
and of sleptons $f_{\widetilde \ell}(\mathbf k)-\bar f_{\widetilde \ell}(\mathbf k)$.
The network of kinetic equations can be expressed as
\begin{subequations}
\label{kin:eq}
\begin{align}
\frac d{d\eta}
\left(
f_\ell(\mathbf k)-\bar f_\ell(\mathbf k)
\right)
&={\cal C}_\ell(\mathbf k)
=\int\frac{dk^0}{2\pi}
{\rm tr}
\left[
{\rm i}\slashed\Sigma^>_\ell(k){\rm i}S_\ell^<(k)
-{\rm i}\slashed\Sigma_\ell^<(k){\rm i}S^>_\ell(k)
\right]\,,
\\
\frac d{d\eta}
\left(
f_{\widetilde \ell}(\mathbf k)-\bar f_{\widetilde \ell}(\mathbf k)
\right)
&={\cal C}_{\widetilde \ell}(\mathbf k)
=-\int\frac{dk^0}{2\pi}
\left[
{\rm i}\Pi^>_{\widetilde \ell}(k){\rm i}\Delta_{\widetilde \ell}^<(k)
-{\rm i}\Pi_{\widetilde \ell}^<(k){\rm i}\Delta^>_{\widetilde \ell}(k)
\right]\,,
\\
\frac d{d\eta}
f_N(\mathbf k)&=
{\cal C}_N(\mathbf k)=
\frac 14\int\frac{dk^0}{2\pi}{\rm sign}(k^0)
{\rm tr}
\left[
{\rm i}\slashed\Sigma^>_N(k){\rm i}S_N^<(k)
-{\rm i}\slashed\Sigma_N^<(k){\rm i}S^>_N(k)
\right]\,,
\end{align}
\end{subequations}
where $\eta$ is the conformal time and $k$ denotes the conformal momentum.
The terms $\cal C$ are referred to as the collision terms. The propagators ${\rm i}\Delta$ (bosons), ${\rm i}S$ (fermions) and the distribution
functions are introduced
in Appendix~\ref{appendix:propagators}.
The expansion of the Universe is then implicitly taken into account when
obtaining the on-shell conformal four-momentum of a particle of mass $m$
through the
relation $k^0=\pm\sqrt{\mathbf k^2+a^2m^2}$, where $a$ is the scale-factor
(see Ref.~\cite{Beneke:2010wd}
for more details). For 
present purposes, it is useful to decompose the self-energies as follows:
\begin{subequations}
\begin{align}
{\rm i}\slashed\Sigma_\ell&=
{\rm i}\slashed\Sigma^{\rm l}_\ell
+{\rm i}\slashed\Sigma^{\rm v}_\ell
+{\rm i}\slashed\Sigma^{\rm w}_\ell+\ldots
\,,
\\
{\rm i}\Pi_\ell&=
{\rm i}\Pi^{\rm l}_{\widetilde \ell}
+{\rm i}\Pi^{\rm v}_{\widetilde \ell}
+{\rm i}\Pi^{\rm w}_{\widetilde \ell}+\ldots
\,,
\\
{\rm i}\slashed\Sigma_N&=
{\rm i}\slashed\Sigma^{\rm l}_N
\,,
\end{align}
\end{subequations}
where the superscript \lq\lq ${\rm l}$" indicates the leading order contributions that
result from one-loop diagrams,
\lq\lq ${\rm v}$" indicates the leading CP-violating contributions from vertex corrections,
\lq\lq ${\rm w}$" indicates the leading CP-violating contributions from wave-function corrections,
that both result from two-loop diagrams, and
the ellipses represent higher order terms that are irrelevant for
the calculation of the lepton asymmetry. In an analogous manner, we also
decompose the collision terms into ${\cal C}_\ell^{\rm l,v,w}$ and
${\cal C}_{\widetilde \ell}^{\rm l,v,w}$.

Note that obtaining the solution to Eqs.~(\ref{kin:eq})
is not the aim of this paper. They are presented here
in order to illustrate the context in which the main objects of our scrutiny -- 
the CPV contributions to the collision terms -- occur.
For the sake of completeness and setting notation,
we list the leading, CP-conserving terms
\begin{subequations}
\begin{align}
{\rm i}\Pi_{\widetilde \ell}^{{\rm l}ab}(k)
=&-Y^2\int\frac{d^4p}{(2\pi)^4}
{\rm tr}\left[
{\rm i}S_N^{ab}(p+k)P_{\rm L}{\rm i}S^{ba}_{\widetilde H}(p)P_{\rm R}
\right]
\\\notag
-&\frac{g_1^2}{2}\int\frac{d^4p}{(2\pi)^4}
{\rm tr}\left[
{\rm i}S_{\ell}^{ab}(p+k)P_{\rm R}{\rm i}S_{\widetilde B}(-p)P_{\rm L}
\right]
\,,
\\
{\rm i}\slashed\Sigma_{\ell}^{{\rm l}ab}(k)
=&
Y^2\int\frac{d^4p}{(2\pi)^4}
P_{\rm R}{\rm i}S_N^{ab}(p+k)P_{\rm L}{\rm i}\Delta^{ba}_H(p)
\\\notag
+&\frac{g_1^2}{2}\int\frac{d^4p}{(2\pi)^4}
{\rm i}\Delta_{\widetilde \ell}^{ab}(p+k)P_{\rm R}{\rm i}S_{\widetilde B}^{ab}(-p)P_{\rm L}\,,
\\\notag
{\rm i}\slashed\Sigma_{N}^{{\rm l}ab}(k)
=&Y^2\int\frac{d^4p}{(2\pi)^4}
\Big[
P_{\rm L}{\rm i}S_\ell^{ab}(p+k)P_{\rm R}{\rm i}\Delta^{ab}_H(-p)
+C\left[P_{\rm L}{\rm i}S_\ell^{ba}(-p-k)P_{\rm R}\right]^TC^\dagger{\rm i}\Delta^{ba}_H(p)
\\
&\hskip1cm+
{\rm i}\Delta_{\widetilde \ell}^{ab}(p+k)
P_{\rm L}{\rm i}S_{\widetilde H}^{ab}(-p)P_{\rm R}
+{\rm i}\Delta_{\widetilde \ell}^{ba}(-p-k)
C\left[P_{\rm L}{\rm i}S_{\widetilde H}^{ba}(p)P_{\rm R}\right]^TC^\dagger
\Big]
\,.
\end{align}
\end{subequations}
The superscripts  $a$, $b$, {\em etc.} take on values +1 or -1, depending on whether the originating or terminating vertex lies on the
forward or backward going branch of the closed time path, respectively. 

\section{Cancellation of Contributions from Vertex Diagrams}
\label{sec:Vertex}

We now proceed to demonstrate the vanishing of the CPV lepton number asymmetry, organizing the discussion according to different diagram topologies. One may map the latter onto the amplitudes entering conventional asymmetry calculations according to the relevant cuts. The first class, which we denote as \lq\lq vertex-type" self-energies, correspond to asymmetry contributions generated by the interference of tree-level and one-loop vertex correction  amplitudes. The second class, analyzed in Section \ref{section:wv}, are equivalent to the interference of tree-level and one-loop wavefunction correction graphs. 

\subsection{Vertex-Type Self-Energies in the CTP}

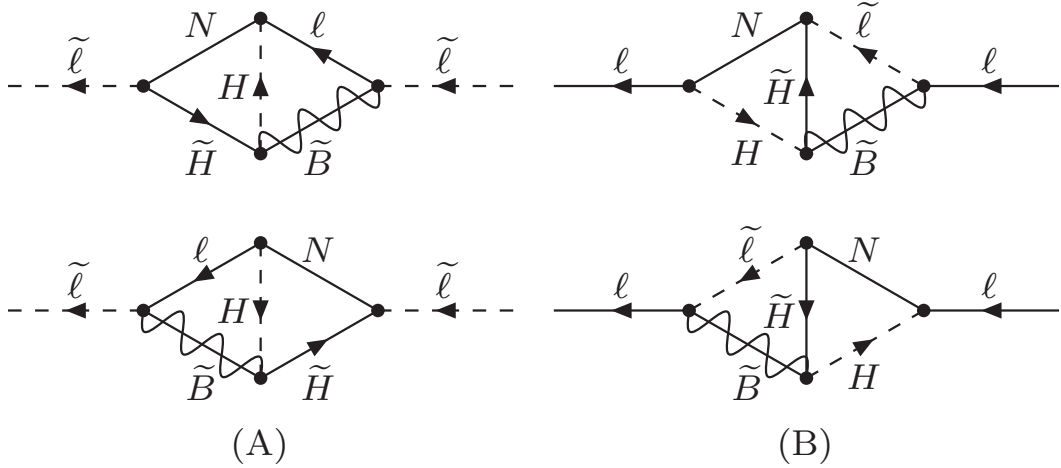
\begin{figure}[ht!]
\begin{center}
\scalebox{1.7}{
\begin{picture}(112,100)(0,0)

\DashArrowLine(30,30)(0,30){3}
\ArrowLine(56,15)(82,30)
\Line(30,30)(56,15)
\Photon(56,15)(30,30){-4}{3}
\Line(82,30)(56,45)
\ArrowLine(56,45)(30,30)
\DashArrowLine(112,30)(82,30){3}
\DashArrowLine(56,45)(56,15){3}
\Vertex(30,30){1.5}
\Vertex(56,15){1.5}
\Vertex(56,45){1.5}
\Vertex(82,30){1.5}
\Text(56,0)[]{\scriptsize{(A)}}
\Text(15,33)[b]{{$\scriptstyle\widetilde \ell$}}
\Text(97,33)[b]{{$\scriptstyle\widetilde \ell$}}
\Text(69,41)[b]{{$\scriptstyle N$}}
\Text(43,41)[b]{{$\scriptstyle \ell$}}
\Text(69,18)[t]{{$\scriptstyle\widetilde H$}}
\Text(43,18)[t]{{$\scriptstyle \widetilde B$}}
\Text(54,30)[r]{{$\scriptstyle H$}}

\SetOffset(0,50)
\DashArrowLine(30,30)(0,30){3}
\ArrowLine(30,30)(56,15)
\Line(82,30)(56,15)
\Photon(82,30)(56,15){4}{3}
\Line(30,30)(56,45)
\ArrowLine(82,30)(56,45)
\DashArrowLine(112,30)(82,30){3}
\DashArrowLine(56,15)(56,45){3}
\Vertex(30,30){1.5}
\Vertex(56,15){1.5}
\Vertex(56,45){1.5}
\Vertex(82,30){1.5}
\Text(15,33)[b]{{$\scriptstyle\widetilde \ell$}}
\Text(97,33)[b]{{$\scriptstyle\widetilde \ell$}}
\Text(43,41)[b]{{$\scriptstyle N$}}
\Text(69,41)[b]{{$\scriptstyle \ell$}}
\Text(43,18)[t]{{$\scriptstyle\widetilde H$}}
\Text(69,18)[t]{{$\scriptstyle \widetilde B$}}
\Text(54,30)[r]{{$\scriptstyle H$}}
\end{picture}
}
\scalebox{1.7}{
\begin{picture}(112,100)(0,0)

\ArrowLine(30,30)(0,30)
\DashArrowLine(56,15)(82,30){3}
\Line(30,30)(56,15)
\Photon(56,15)(30,30){-4}{3}
\Line(82,30)(56,45)
\DashArrowLine(56,45)(30,30){3}
\ArrowLine(112,30)(82,30)
\ArrowLine(56,45)(56,15)
\Vertex(30,30){1.5}
\Vertex(56,15){1.5}
\Vertex(56,45){1.5}
\Vertex(82,30){1.5}
\Text(56,0)[]{\scriptsize{(B)}}
\Text(15,33)[b]{{$\scriptstyle \ell$}}
\Text(97,33)[b]{{$\scriptstyle \ell$}}
\Text(69,41)[b]{{$\scriptstyle N$}}
\Text(43,41)[b]{{$\scriptstyle\widetilde\ell$}}
\Text(69,18)[t]{{$\scriptstyle H$}}
\Text(43,18)[t]{{$\scriptstyle \widetilde B$}}
\Text(54,30)[r]{{$\scriptstyle\widetilde H$}}

\SetOffset(0,50)
\ArrowLine(30,30)(0,30)
\DashArrowLine(30,30)(56,15){3}
\Line(82,30)(56,15)
\Photon(82,30)(56,15){4}{3}
\Line(30,30)(56,45)
\DashArrowLine(82,30)(56,45){3}
\ArrowLine(112,30)(82,30)
\ArrowLine(56,15)(56,45)
\Vertex(30,30){1.5}
\Vertex(56,15){1.5}
\Vertex(56,45){1.5}
\Vertex(82,30){1.5}
\Text(15,33)[b]{{$\scriptstyle\ell$}}
\Text(97,33)[b]{{$\scriptstyle\ell$}}
\Text(43,41)[b]{{$\scriptstyle N$}}
\Text(69,41)[b]{{$\scriptstyle\widetilde\ell$}}
\Text(43,18)[t]{{$\scriptstyle H$}}
\Text(69,18)[t]{{$\scriptstyle \widetilde B$}}
\Text(54,30)[r]{{$\scriptstyle\widetilde H$}}
\end{picture}
}
\end{center}
\caption{\label{fig:selfergs}The vertex-type self-energies
${\rm i}\Pi_{\widetilde \ell}^{\rm v}$~(A) and ${\rm i}\slashed\Sigma^{\rm v}_\ell$~(B).}
\end{figure}

The relevant vertex-type self-energies that lead to source terms of asymmetries in
$\ell$ and $\widetilde \ell$ are represented diagrammatically in
Fig.~\ref{fig:selfergs}. We evaluate these and extract the
contributions that are important when $\ell$, $\widetilde \ell$,
$H_{1,2}$, $\Psi_{\widetilde H^{\pm}}$ and $\Psi_{\widetilde B}$
are in equilibrium following Ref.~\cite{Beneke:2010wd}.

The diagrams in Fig.~\ref{fig:selfergs}(A) correspond to the CTP self-energy for the slepton
$\widetilde \ell$
\begin{align}
\label{Pi:ltilde}
{\rm i}\Pi^{{\rm v}ab}_{\widetilde \ell}(k)=&
-cd Y^2 \frac{g_1^2}{2}{\rm tr}
\int\frac{d^4p}{(2\pi)^4}\frac{d^4q}{(2\pi)^4}
\big\{
\\\notag
&
P_{\rm L}{\rm i}S_\ell^{ac}(-p)
{\rm i}S_N^{cb}(q+k)
P_{\rm L}
{\rm i} S^{bd}_{\widetilde H}(q)
P_{\rm L}{\rm i}S^{da}_{\widetilde B}(-p-k)
{\rm i}\Delta_{H1}^{dc}(p+k+q)
\sin\alpha \cos\alpha{\rm e}^{-{\rm i}\phi_\mu}
\\\notag
+&
P_{\rm R}{\rm i}S_N^{ac}(-p)
{\rm i}S_{\ell}^{cb}(q+k)
P_{\rm R}
{\rm i}S^{bd}_{\widetilde B}(q)
P_{\rm R}
{\rm i} S_{\widetilde H}^{da}(-p-k)
{\rm i}\Delta^{cd}_{H1}(-p-k-q)
\sin\alpha \cos\alpha{\rm e}^{{\rm i}\phi_\mu}
\big\}\,.
\end{align}
For simplicity, we assume here and in the following
that $m_{H2}\gg m_{H1}$
and $m_{H2}\gg T$, such that
contributions form an internal line of $H_2$ may be neglected.
The more general expressions that also include $H_2$ propagators
can easily be inferred from the results presented here.
Furthermore,
we have only accounted for those contributions that lead to CP-violation
(i.e. we have neglected additional terms originating from the mixing of
$H_{1,2}$ which do not lead to CP-violation).

The relevant contributions
to the self-energy of the lepton $\ell$
are depicted in Fig.~\ref{fig:selfergs}(B).
Again, we keep only the CPV contribution that is mediated by the lighter Higgs boson
$H_1$. We find
\begin{align}
\label{Sigma:ell}
{\rm i}{\Sigma\!\!\!\!/}_\ell^{{\rm v}ab}(k)=&cd Y^2 \frac{g_1^2}{2}
\int \frac{d^4p}{(2\pi)^4}\frac{d^4 q}{(2\pi)^4}\Big\{
\\\notag
&
P_{\rm R}{\rm i} S_{\widetilde B}^{ad}(p+k) P_{\rm R} {\rm i} S_{\widetilde H}^{dc}(p+k+q)
P_{\rm R}{\rm i} S_N^{cb}(q+k)P_{\rm L} {\rm i} \Delta_{\widetilde \ell}^{ac}(-p)
{\rm i}\Delta_{H1}^{bd}(q)
\sin\alpha\cos\alpha {\rm e}^{{\rm i}\phi_\mu}
\\\notag
+&
P_{\rm R}{\rm i} S_{N}^{ac}(-p) P_{\rm L} {\rm i} S_{\widetilde H}^{cd}(-p-k-q)
P_{\rm L}{\rm i} S_{\widetilde B}^{db}(-q) P_{\rm L} {\rm i} \Delta_{\widetilde \ell}^{cb}(q+k)
{\rm i}\Delta_{H1}^{da}(-p-k)
\sin\alpha\cos\alpha {\rm e}^{-{\rm i}\phi_\mu}
\Big\}\,.
\end{align}

\subsection{Extracting the CPV Contributions}

For purposes of simplicity, we assume that
those particles charged under the electroweak gauge group are in kinetic equilibrium.
In order to extract the leading contributions to the asymmetries
in leptons $\ell$ and sleptons $\widetilde \ell$, we substitute
equilibrium distributions into the CPV sources
${\cal C^{\rm v,w}}$: $f_B (\mathbf k)= 1/[ \exp(E/T) -1]$ and $f_F (\mathbf k)= 1/[ \exp(E/T) +1]$ with $E=\sqrt{\mathbf k^2+m^2}$
for the bosons $B=\widetilde \ell,H1,H2$ and fermions $F=\ell,\widetilde H,\widetilde B$.
For the bosonic and fermionic
equilibrium propagators, one may also apply the Kubo-Martin-Schwinger (KMS) relations
\begin{subequations}
\begin{align}
{\rm i}\Delta^>(p)&={\rm e}^{p^0/T}{\rm i}\Delta^<(p)\,,
\\
{\rm i}S^>(p)&=-{\rm e}^{p^0/T}{\rm i}S^<(p)\,,
\end{align}
\end{subequations}
where for a general Green's function $G^{ab}$ one has $G^{-+}\equiv G^>$ and $G^{-+}\equiv G^<$.
The singlet neutrino $N$ is in general out-of-equilibrium.
We sometimes denote by $\delta S_N$ or $\delta f_N$ the difference
between the propagator or the distribution function of $N$
and their equilibrium counterparts.
We note in passing that in the terms ${\cal C}^{\rm l}$, that describe the
washout of the charges, distributions with chemical potential
instead of the above equilibrium
distributions should be substituted~\cite{Beneke:2010wd}.

In order to calculate the leading CPV contribution from vertex diagrams,
${\cal C}^{\rm v}$, we may substitute
tree-level equilibrium propagators
for ${\rm i}S_\ell$, ${\rm i}\Delta_{\widetilde \ell}$,
${\rm i}S_{\widetilde B}$ and ${\rm i}\Delta_{H}$, while allowing
for a non-equilibrium form for ${\rm i}S_N$.
From Fig.~\ref{fig:selfergs}
and from the expressions~(\ref{Pi:ltilde}) and~(\ref{Sigma:ell}) for the self-energies, we
see that for both leptons and sleptons,
there occur two diagrammatic contributions to the collision terms that
are related by complex conjugation and a reversed fermion flow.
(Note that ${\cal C}_{\ell,\widetilde \ell}^{\rm v}$ may be interpreted
as being a result of closing the external lines in the diagrams of
Fig.~\ref{fig:selfergs}.)
In order to obtain a result that depends on the CPV phases,
at least one of the internal propagators must be imaginary, i.e. off-shell.
It is therefore suitable to distinguish the various contributions
according to which of the particles is taken to be off-shell, writing
${\cal C}^{\rm v}_{\widetilde \ell}={\cal C}^{{\rm v}\widetilde B}_{\widetilde \ell}+{\cal C}^{{\rm v}\widetilde H}_{\widetilde \ell}+{\cal C}^{{\rm v} H}_{\widetilde \ell}+{\cal C}^{{\rm v} \ell}_{\widetilde \ell}$ and
${\cal C}^{\rm v}_{\ell}={\cal C}^{{\rm v}\widetilde B}_{\ell}+{\cal C}^{{\rm v}\widetilde H}_{\ell}+{\cal C}^{{\rm v} H}_{\ell}+{\cal C}^{{\rm v} \widetilde\ell}_{\ell}$ with the off-shell particle in each component indicated by the superscript. 
This
procedure can also be understood from considering the corresponding
diagrams for vacuum decay and their cuts that we provide for each case.

With these remarks in mind, we outline a procedure for extracting the CPV 
contributions from the self-energies~(\ref{Pi:ltilde})
and~(\ref{Sigma:ell}). Essentially, it
is the same method that is applied to conventional leptogenesis
in Ref.~\cite{Beneke:2010wd}:
\begin{itemize}
\item
Write down all terms that contribute to $\Sigma^>$ or $\Pi^>$. The $<$ terms follow when replacing all CTP indices $+\leftrightarrow -$.

\item
Take one particular propagator off-shell, such that only the (anti)-time ordered components contribute.

\item
Write down the collision terms ${\cal C}^{\rm v}$
or ${\cal C}^{\rm wv}$. There are eight individual terms (for both,
vertex and wave-function contributions).

\item
Make use of the relations between the two-point functions, e.g. $G^T+G^{\bar T}= G^>+G^<$ is useful. Other identities, that hold under the integrals and will be used for the vertex diagrams are presented in Ref.~\cite{Beneke:2010wd}. (See, for example,
relation~(\ref{repl:TTBAR:GRLESS}) below.)

\item
Use KMS relations in order to establish the cancellation or the vanishing of the particular contributions.

\item
 Make use of the behavior of the $G^{ab}$ under momentum reversals, while assuming spatial isotropy of the collision terms:
${\cal C}_{\ell}(\mathbf k)={\cal C}_{\ell}(-\mathbf k)$ {\em etc.}
\end{itemize}

\subsection{Cuts with off-shell $\widetilde B$}
\label{sec:offshellbino}

\begin{figure}[ht!]
\begin{center}
\scalebox{1.7}{
\begin{picture}(86,60)(0,0)

\Line(0,30)(30,30)
\Line(30,30)(56,15)
\DashLine(30,30)(56,45){3}
\Line(56,15)(56,45)
\Photon(56,15)(56,45){4}{3}
\Line(56,45)(86,45)
\DashLine(56,15)(86,15){3}
\Vertex(30,30){1.5}
\Vertex(56,15){1.5}
\Vertex(56,45){1.5}
\Text(15,33)[b]{{$\scriptstyle N$}}
\Text(43,41)[b]{{$\scriptstyle \widetilde\ell$}}
\Text(71,12)[t]{{$\scriptstyle H$}}
\Text(71,47)[b]{{$\scriptstyle \ell$}}
\Text(43,18)[t]{{$\scriptstyle \widetilde H$}}
\Text(52,30)[r]{{$\scriptstyle \widetilde B$}}

\SetColor{Red}
\Line(35,20)(35,40)
\Line(36,20)(36,40)

\end{picture}
}
\scalebox{1.7}{
\begin{picture}(86,60)(0,0)

\Line(0,30)(30,30)
\DashLine(30,30)(56,15){3}
\Line(30,30)(56,45)
\Line(56,15)(56,45)
\Photon(56,15)(56,45){4}{3}
\DashLine(56,45)(86,45){3}
\Line(56,15)(86,15)
\Vertex(30,30){1.5}
\Vertex(56,15){1.5}
\Vertex(56,45){1.5}
\Text(15,33)[b]{{$\scriptstyle N$}}
\Text(43,41)[b]{{$\scriptstyle \ell$}}
\Text(71,12)[t]{{$\scriptstyle \widetilde H$}}
\Text(71,47)[b]{{$\scriptstyle \widetilde\ell$}}
\Text(43,18)[t]{{$\scriptstyle H$}}
\Text(52,30)[r]{{$\scriptstyle \widetilde B$}}

\SetColor{Red}
\Line(35,20)(35,40)
\Line(36,20)(36,40)

\end{picture}
}
\end{center}
\caption{\label{fig:offbino}Cuts in vacuum diagrams with off-shell bino.}
\end{figure}
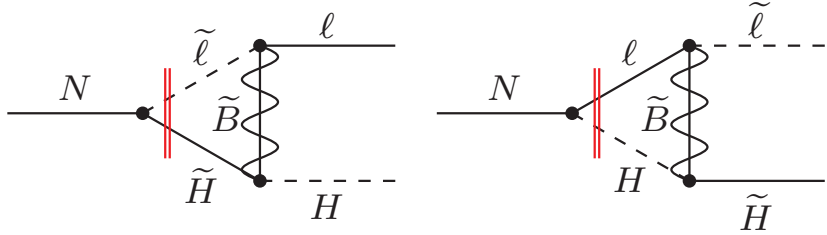

 In this Section, we take the bino $\widetilde B$ to be off-shell. The correspondence with the 
Boltzmann approach can been seen from the appropriate cuts in the diagrams of Fig.~\ref{fig:selfergs}. First cut the graphs through the $N$ line and either the  ${\widetilde H}$ line in Fig.~\ref{fig:selfergs} (A) or the $H$ line in Fig.~\ref{fig:selfergs} (B). The result is a contribution to the asymmetries for a final state  (${\widetilde\ell}$, ${\widetilde H}$) or ($\ell$, $H_1$) pair, respectively, corresponding to the interference of the tree-level and one-loop vertex correction amplitudes. The off-shell bino contribution then corresponds to the additional cuts in the vertex correction graphs shown in Fig.~\ref{fig:offbino}. 
For this type of contribution, ${\cal C}_{\widetilde B}$,
it has been proposed that
when including thermal corrections, there will be a residual lepton 
asymmetry~\cite{Fong:2009iu}. Instead of decays of the singlet fermion
$N$, decays of its superpartner $\widetilde N$ are discussed in
Ref.~\cite{Fong:2009iu}. The considerations that are presented here
can however be transferred to the decays of scalars straightforwardly.

In the vacuum,
the asymmetries from decays of $N$ that are stored in leptons
and sleptons are precisely opposite. When calculating the
CPV contributions that result from interference of the
loop amplitudes in Fig.~\ref{fig:offbino} with the
tree-level amplitudes by using Cutkosky's rules, we see that
no difference between
the integration over internal on-shell particles
$\widetilde \ell$, $\widetilde H$ or $\ell$, $H$ and the
external phase space integral is made.

Now consider the decays and inverse decays in the finite temperature
background of the Early Universe. It is stated in Ref.~\cite{Fong:2009iu},
that the quantum statistics of the external phase space renders the
aforementioned cancellation ineffective. In the following, we show that
the cancellation is reinstated once the statistical corrections for
the internal particles are accounted for as well.
As explained above, we assume that all particle number distributions except for
the distribution of the right-handed neutrinos $N$ take the equilibrium form.
This allows us to frequently apply KMS relations for all propagators except 
for $S^{<,>}_N$.
A calculation that is very analogous to the one in Ref.~\cite{Beneke:2010wd}
leads to the CPV contribution to the collision term
\begin{align}
\label{Coll:ltilde}
{\cal C}^{{\rm v}\widetilde B}_{\widetilde \ell}(\mathbf k)
=&
Y^2\frac{g_1^2}{4}{\rm tr}
\int\frac{dk^0}{2\pi}\frac{d^4 p}{(2\pi)^4}\frac{d^4q}{(2\pi)^4}
\Big\{
\\\notag
&
\left[
{\rm i}\Delta_{H1}^<(p+k+q)
{\rm i}S_{\ell}^<(-p)
-{\rm i}\Delta_{H1}^>(p+k+q)
{\rm i}S_{\ell}^>(-p)
\right]
P_{\rm R}{\rm i}\delta S_N(q+k)P_{\rm L}
\\\notag
&\times
\left[
{\rm i}S^<_{\widetilde H}(q)
{\rm i}\Delta_{\widetilde \ell}^<(k)
-
{\rm i}S^>_{\widetilde H}(q)
{\rm i}\Delta_{\widetilde \ell}^>(k)
\right]
P_{\rm L}S^T_{\widetilde B}(-p-k)
\sin\alpha\cos\alpha{\rm e}^{-{\rm i}\phi_\mu}
\\\notag
-&
P_{\rm R}{\rm i}\delta S_N(q+k)
\left[
{\rm i}\Delta_{H1}^<(p+k+q)
{\rm i}S_{\ell}^<(-p)
-{\rm i}\Delta_{H1}^>(p+k+q)
{\rm i}S_{\ell}^>(-p)
\right]
\\\notag
&\times
P_{\rm R}S^T_{\widetilde B}(-p-k)P_{\rm R}
\left[
{\rm i}S^<_{\widetilde H}(q)
{\rm i}\Delta_{\widetilde \ell}^<(k)
-
{\rm i}S^>_{\widetilde H}(q)
{\rm i}\Delta_{\widetilde \ell}^>(k)
\right]
\sin\alpha\cos\alpha{\rm e}^{{\rm i}\phi_\mu}
\Big\}\,.
\end{align}
Notice that in transforming expression~(\ref{Pi:ltilde}) to~(\ref{Coll:ltilde}),
we have made the replacements
\begin{subequations}
\label{repl:TTBAR:GRLESS}
\begin{align}
&{\rm i}S_\ell^>(-p){\rm i}\Delta_{H_1}^>(p+k-q)
-{\rm i}S_\ell^{T,\bar T}(-p){\rm i}\Delta_{H_1}^{T,\bar T}(p+k-q)
\\\notag
\to&-\frac12
\left(
{\rm i}S_\ell^<(-p){\rm i}\Delta_{H_1}^<(p+k-q)
-{\rm i}S_\ell^>(-p){\rm i}\Delta_{H_1}^>(p+k-q)
\right)\,,
\\
&{\rm i}S_\ell^<(-p){\rm i}\Delta_{H_1}^<(p+k-q)
-{\rm i}S_\ell^{T,\bar T}(-p){\rm i}\Delta_{H_1}^{T,\bar T}(p+k-q)
\\\notag
\to&+\frac12
\left(
{\rm i}S_\ell^<(-p){\rm i}\Delta_{H_1}^<(p+k-q)
-{\rm i}S_\ell^>(-p){\rm i}\Delta_{H_1}^>(p+k-q)
\right)
\end{align}
\end{subequations}
that do not change the integrals, because the dispersive contributions from
the product of (anti-) time-ordered propagators cancel, as explained
in Appendix~\ref{appendix:replacements} and in
in Ref.~\cite{Beneke:2010wd}.

The collision term~(\ref{Coll:ltilde})
clearly exhibits that in the CTP formalism, there is no distinction between
cut particles and external states, because $\widetilde H$, $\widetilde \ell$ and
$H$, $\ell$ appear in a symmetric way (exchange of the terms in the square brackets). This may be interpreted as a generalization of Cutkosky's
rules from the vacuum background to finite densities.

From the self-energy~(\ref{Sigma:ell}), we find the CPV
contribution to the lepton collision term with off-shell $\widetilde B$,
which is
\begin{align}
{\cal C}_\ell^{{\rm v}\widetilde B}(\mathbf k)=&Y^2 \frac{g_1^2}{4}\sin\alpha\cos\alpha\,{\rm tr}
\int\frac{dk^0}{2\pi}\frac{d^4p}{(2\pi)^4}\frac{d^4q}{(2\pi)^4}
\Big\{
\\\notag
&
{\rm i}S^T_{\widetilde B}(p+k)P_{\rm R}\left[
{\rm i}S_{\widetilde H}^<(p+k+q){\rm i}\Delta_{\widetilde \ell}^<(-p)
-{\rm i}S_{\widetilde H}^>(p+k+q){\rm i}\Delta_{\widetilde \ell}^>(-p)
\right]
P_{\rm R}{\rm i}\delta S_N(q+k)
\\\notag
&\times P_{\rm L}\left[
{\rm i}\Delta_{H1}^<(q){\rm i}S_\ell^<(k)
-
{\rm i}\Delta_{H1}^>(q){\rm i}S_\ell^>(k)
\right]
{\rm e}^{{\rm i}\phi_\mu}
\\\notag
-&
{\rm i}\delta S_N(q+k)P_{\rm L}\left[
{\rm i}S_{\widetilde H}^<(p+k+q){\rm i}\Delta_{\widetilde \ell}^<(-p)
-{\rm i}S_{\widetilde H}^>(p+k+q){\rm i}\Delta_{\widetilde \ell}^>(-p)
\right]
P_{\rm L}
{\rm i}S^T_{\widetilde B}(p+k)
\\\notag
&\times P_{\rm L}\left[
{\rm i}\Delta_{H1}^<(q){\rm i}S_\ell^<(k)
-
{\rm i}\Delta_{H1}^>(q){\rm i}S_\ell^>(k)
\right]
{\rm e}^{-{\rm i}\phi_\mu}
\Big\}
\,.
\end{align}

When relabeling the momentum variables, it is now easy to see that
\begin{align}
\label{eq:cancel1}
\int \frac{d^3k}{(2\pi)^3}{\cal C}^{{\rm v}\widetilde B}_{\widetilde \ell}({\mathbf k})
=-\int \frac{d^3k}{(2\pi)^3}{\cal C}^{{\rm v}\widetilde B}_{\ell}({\mathbf k})\,.
\end{align}

\begin{figure}[t!]
\begin{center}
\scalebox{1.5}{
\begin{picture}(280,236)(0,0)

\SetOffset(0,195)
\Text(270,30)[l]{$\scriptstyle \textnormal{(A)}$}
\SetOffset(65,195)

\ArrowLine(30,30)(0,30)
\DashArrowLine(56,15)(82,30){3}
\Line(30,30)(56,15)
\Photon(56,15)(30,30){-4}{3}
\Line(82,30)(56,45)
\DashArrowLine(56,45)(30,30){3}
\ArrowLine(112,30)(82,30)
\ArrowLine(56,45)(56,15)
\Vertex(30,30){1.5}
\Vertex(56,15){1.5}
\Vertex(56,45){1.5}
\Vertex(82,30){1.5}
\Text(15,33)[b]{{$\scriptstyle \ell$}}
\Text(97,33)[b]{{$\scriptstyle \ell$}}
\Text(69,41)[b]{{$\scriptstyle N$}}
\Text(43,41)[b]{{$\scriptstyle\widetilde\ell$}}
\Text(69,18)[t]{{$\scriptstyle H$}}
\Text(43,18)[t]{{$\scriptstyle \widetilde B$}}
\Text(58,35)[l]{{$\scriptstyle\widetilde H$}}

\SetColor{Red}
\Line(77,20)(77,40)
\Line(76,20)(76,40)
\SetColor{Orange}
\Line(30,36)(75,24)
\Line(30,35)(75,23)
\SetColor{Black}

\SetOffset(0,130)

\Text(120,60)[]{$\Longrightarrow$}

\Text(270,30)[l]{$\scriptstyle \textnormal{(B)}$}

\Line(0,30)(15,30)
\ArrowLine(15,30)(41,15)
\DashArrowLine(15,30)(41,45){3}
\Line(41,15)(41,45)
\Photon(41,15)(41,45){4}{3}
\ArrowLine(41,45)(56,45)
\DashArrowLine(41,15)(56,15){3}
\Vertex(15,30){1.5}
\Vertex(41,15){1.5}
\Vertex(41,45){1.5}
\Text(7,33)[b]{{$\scriptstyle N$}}
\Text(28,41)[b]{{$\scriptstyle \widetilde\ell$}}
\Text(49,12)[t]{{$\scriptstyle H$}}
\Text(49,47)[b]{{$\scriptstyle \ell$}}
\Text(28,18)[t]{{$\scriptstyle \widetilde H$}}
\Text(37,30)[r]{{$\scriptstyle \widetilde B$}}

\Text(63,30)[]{{$\scriptstyle \times$}}
\Text(72,30)[]{$\Bigg($}
\Text(117,30)[]{$\Bigg)$}
\Text(120,45)[]{$\scriptstyle *$}

\Line(74,30)(94,30)
\ArrowLine(94,30)(111,40)
\DashArrowLine(94,30)(111,20){3}
\Vertex(94,30){1.5}
\Text(84,33)[b]{{$\scriptstyle N$}}
\Text(103,40)[b]{{$\scriptstyle \ell$}}
\Text(103,20)[t]{{$\scriptstyle H$}}

\Text(130,30)[]{$\scriptstyle +$}

\SetOffset(147,130)

\ArrowLine(0,13)(17,23)
\DashArrowLine(17,23)(34,13){3}
\Line(17,23)(17,38)
\Photon(17,23)(17,38){3}{2.5}
\DashArrowLine(0,48)(17,38){3}
\ArrowLine(17,38)(34,48)
\Text(1,17)[b]{$\scriptstyle \widetilde H$}
\Text(33,17)[b]{$\scriptstyle H$}
\Text(1,44)[t]{$\scriptstyle \widetilde \ell$}
\Text(33,44)[t]{$\scriptstyle \ell$}
\Text(13,30)[r]{$\scriptstyle \widetilde B$}
\Vertex(17,23){1.5}
\Vertex(17,38){1.5}

\SetOffset(142,130)

\Text(50,30)[]{$\scriptstyle \times$}
\Text(59,30)[]{$\Big($}
\ArrowLine(64,20)(81,30)
\DashArrowLine(64,40)(81,30){3}
\Line(81,30)(96,30)
\ArrowLine(96,30)(113,40)
\DashArrowLine(96,30)(113,20){3}
\Vertex(81,30){1.5}
\Vertex(96,30){1.5}
\Text(118,30)[]{$\Big)$}
\Text(120,40)[]{$\scriptstyle *$}
\Text(104,38)[b]{$\scriptstyle \ell$}
\Text(104,21)[t]{$\scriptstyle H$}
\Text(73,39)[b]{$\scriptstyle \widetilde \ell$}
\Text(73,22)[t]{$\scriptstyle \widetilde H$}
\Text(88,33)[b]{$\scriptstyle N$}


\SetOffset(0,65)
\Line(15,60)(245,60)

\Text(270,30)[l]{$\scriptstyle \textnormal{(A)}^*$}
\SetOffset(65,65)

\ArrowLine(30,30)(0,30)
\DashArrowLine(30,30)(56,15){3}
\Line(56,15)(82,30)
\Photon(56,15)(82,30){-4}{3}
\Line(30,30)(56,45)
\DashArrowLine(82,30)(56,45){3}
\ArrowLine(112,30)(82,30)
\ArrowLine(56,15)(56,45)
\Vertex(30,30){1.5}
\Vertex(56,15){1.5}
\Vertex(56,45){1.5}
\Vertex(82,30){1.5}
\Text(15,33)[b]{{$\scriptstyle \ell$}}
\Text(97,33)[b]{{$\scriptstyle \ell$}}
\Text(69,41)[b]{{$\scriptstyle \widetilde\ell$}}
\Text(43,41)[b]{{$\scriptstyle N$}}
\Text(69,18)[t]{{$\scriptstyle \widetilde B$}}
\Text(43,18)[t]{{$\scriptstyle H$}}
\Text(54,36)[r]{{$\scriptstyle\widetilde H$}}

\SetColor{Red}
\Line(34,20)(34,40)
\Line(35,20)(35,40)
\SetColor{Orange}
\Line(36,25)(82,38)
\Line(36,26)(82,39)
\SetColor{Black}

\SetOffset(0,0)

\Text(120,60)[]{$\Longrightarrow$}

\Text(270,30)[l]{$\scriptstyle \textnormal{(B)}^*$}

\SetOffset(2,0)

\Line(0,30)(15,30)
\ArrowLine(15,30)(41,15)
\DashArrowLine(15,30)(41,45){3}
\Line(41,15)(41,45)
\Photon(41,15)(41,45){4}{3}
\ArrowLine(41,45)(56,45)
\DashArrowLine(41,15)(56,15){3}
\Vertex(15,30){1.5}
\Vertex(41,15){1.5}
\Vertex(41,45){1.5}
\Text(7,33)[b]{{$\scriptstyle N$}}
\Text(28,41)[b]{{$\scriptstyle \widetilde\ell$}}
\Text(49,12)[t]{{$\scriptstyle H$}}
\Text(49,47)[b]{{$\scriptstyle \ell$}}
\Text(28,18)[t]{{$\scriptstyle \widetilde H$}}
\Text(37,30)[r]{{$\scriptstyle \widetilde B$}}

\SetOffset(0,0)

\Text(72,30)[]{{$\scriptstyle \times$}}
\Text(0,30)[]{$\Bigg($}
\Text(64,30)[]{$\Bigg)$}
\Text(67,45)[]{$\scriptstyle *$}

\SetOffset(4,0)

\Line(74,30)(94,30)
\ArrowLine(94,30)(111,40)
\DashArrowLine(94,30)(111,20){3}
\Vertex(94,30){1.5}
\Text(84,33)[b]{{$\scriptstyle N$}}
\Text(103,40)[b]{{$\scriptstyle \ell$}}
\Text(103,20)[t]{{$\scriptstyle H$}}

\SetOffset(0,0)

\Text(127,30)[]{$\scriptstyle +$}

\SetOffset(147,0)

\ArrowLine(0,13)(17,23)
\DashArrowLine(17,23)(34,13){3}
\Line(17,23)(17,38)
\Photon(17,23)(17,38){3}{2.5}
\DashArrowLine(0,48)(17,38){3}
\ArrowLine(17,38)(34,48)
\Text(1,17)[b]{$\scriptstyle \widetilde H$}
\Text(33,17)[b]{$\scriptstyle H$}
\Text(1,44)[t]{$\scriptstyle \widetilde \ell$}
\Text(33,44)[t]{$\scriptstyle \ell$}
\Text(13,30)[r]{$\scriptstyle \widetilde B$}
\Vertex(17,23){1.5}
\Vertex(17,38){1.5}

\Text(-8,30)[]{$\Bigg($}
\Text(42,30)[]{$\Bigg)$}
\Text(45,45)[]{$\scriptstyle *$}

\Text(54,30)[]{$\scriptstyle \times$}

\SetOffset(144,0)

\ArrowLine(64,20)(81,30)
\DashArrowLine(64,40)(81,30){3}
\Line(81,30)(96,30)
\ArrowLine(96,30)(113,40)
\DashArrowLine(96,30)(113,20){3}
\Vertex(81,30){1.5}
\Vertex(96,30){1.5}
\Text(104,38)[b]{$\scriptstyle \ell$}
\Text(104,21)[t]{$\scriptstyle H$}
\Text(73,39)[b]{$\scriptstyle \widetilde \ell$}
\Text(73,22)[t]{$\scriptstyle \widetilde H$}
\Text(88,33)[b]{$\scriptstyle N$}

\end{picture}
}
\end{center}
\caption{\label{fig:vertex:cuts} Relation of the lepton CTP self-energies
$\textnormal{(A)}$, $\textnormal{(A)}^*$
to the CPV interference of amplitudes
$\textnormal{(B)}$, $\textnormal{(B)}^*$
in conventional approaches,
on the example of the contribution from the off-shell $\widetilde B$,
where the star indicates that the diagrams represent terms that
are related by complex conjugation.
The CPV contributions to the interference terms occur when
$\widetilde \ell$ and $\widetilde H$ are on-shell. The RIS contribution
arises from an on-shell $N$ in the $2\leftrightarrow 2$ scattering amplitude. In the interference terms, we have suppressed the notation of
the integration over the phase space of the external particles other
than $\ell$.}
\end{figure}
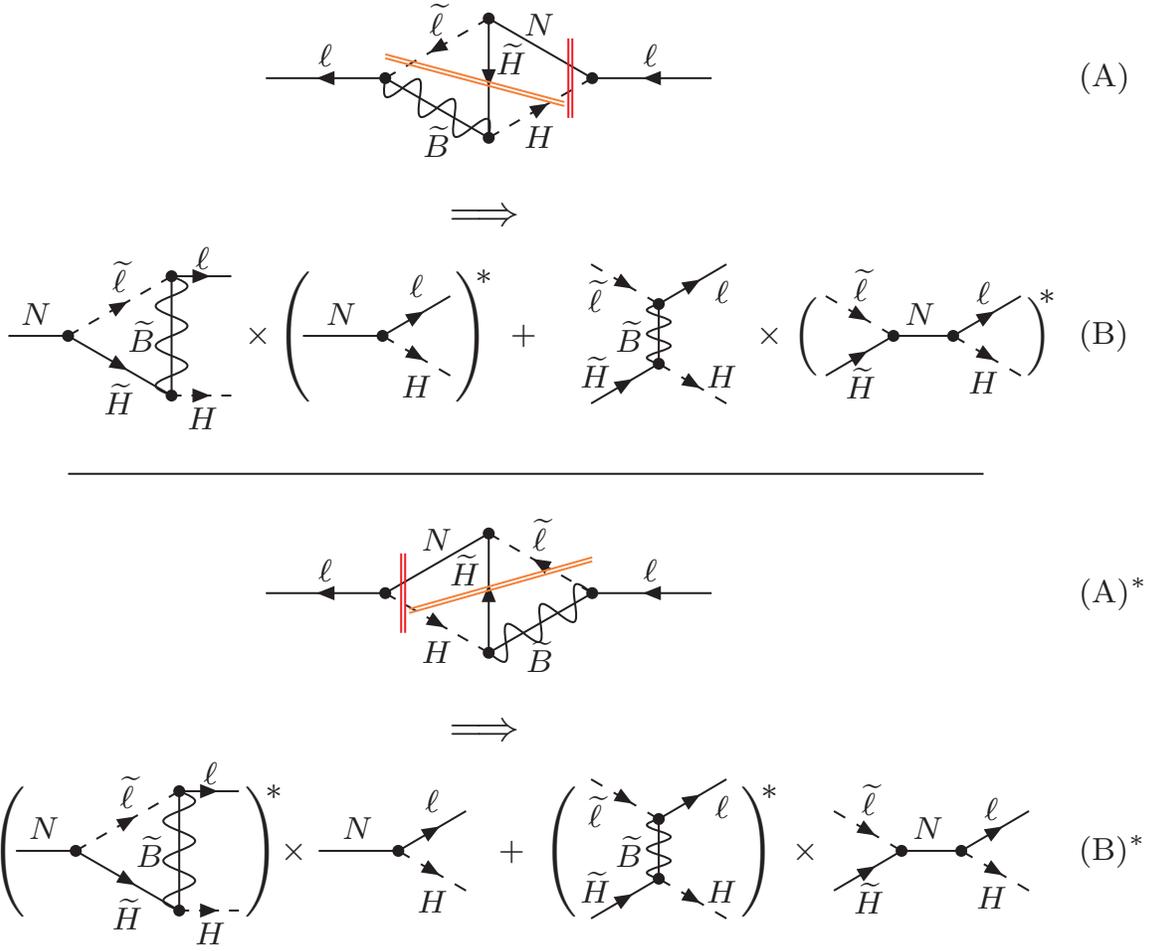

\begin{figure}[t!]
\begin{center}
\scalebox{1.5}{
\begin{picture}(280,236)(0,0)

\SetOffset(0,195)
\Text(270,30)[l]{$\scriptstyle \textnormal{(A)}$}
\SetOffset(65,195)

\DashArrowLine(30,30)(0,30){3}
\ArrowLine(56,15)(82,30)
\Line(30,30)(56,15)
\Photon(56,15)(30,30){-4}{3}
\Line(82,30)(56,45)
\ArrowLine(56,45)(30,30)
\DashArrowLine(112,30)(82,30){3}
\DashArrowLine(56,45)(56,15){3}
\Vertex(30,30){1.5}
\Vertex(56,15){1.5}
\Vertex(56,45){1.5}
\Vertex(82,30){1.5}
\Text(15,33)[b]{{$\scriptstyle \widetilde\ell$}}
\Text(97,33)[b]{{$\scriptstyle \widetilde\ell$}}
\Text(69,41)[b]{{$\scriptstyle N$}}
\Text(43,41)[b]{{$\scriptstyle\ell$}}
\Text(69,18)[t]{{$\scriptstyle \widetilde H$}}
\Text(43,18)[t]{{$\scriptstyle \widetilde B$}}
\Text(58,35)[l]{{$\scriptstyle H$}}

\SetColor{Red}
\Line(77,20)(77,40)
\Line(76,20)(76,40)
\SetColor{Orange}
\Line(30,36)(75,24)
\Line(30,35)(75,23)
\SetColor{Black}

\SetOffset(0,130)

\Text(120,60)[]{$\Longrightarrow$}

\Text(270,30)[l]{$\scriptstyle \textnormal{(B)}$}

\Line(0,30)(15,30)
\DashArrowLine(15,30)(41,15){3}
\ArrowLine(15,30)(41,45)
\Line(41,15)(41,45)
\Photon(41,15)(41,45){4}{3}
\DashArrowLine(41,45)(56,45){3}
\ArrowLine(41,15)(56,15)
\Vertex(15,30){1.5}
\Vertex(41,15){1.5}
\Vertex(41,45){1.5}
\Text(7,33)[b]{{$\scriptstyle N$}}
\Text(28,41)[b]{{$\scriptstyle \ell$}}
\Text(49,12)[t]{{$\scriptstyle \widetilde H$}}
\Text(49,47)[b]{{$\scriptstyle \widetilde\ell$}}
\Text(28,18)[t]{{$\scriptstyle H$}}
\Text(37,30)[r]{{$\scriptstyle \widetilde B$}}

\Text(63,30)[]{{$\scriptstyle \times$}}
\Text(72,30)[]{$\Bigg($}
\Text(117,30)[]{$\Bigg)$}
\Text(120,45)[]{$\scriptstyle *$}

\Line(74,30)(94,30)
\DashArrowLine(94,30)(111,40){3}
\ArrowLine(94,30)(111,20)
\Vertex(94,30){1.5}
\Text(84,33)[b]{{$\scriptstyle N$}}
\Text(103,40)[b]{{$\scriptstyle \widetilde \ell$}}
\Text(103,20)[t]{{$\scriptstyle \widetilde H$}}

\Text(130,30)[]{$\scriptstyle +$}

\SetOffset(147,130)

\DashArrowLine(0,13)(17,23){3}
\ArrowLine(17,23)(34,13)
\Line(17,23)(17,38)
\Photon(17,23)(17,38){3}{2.5}
\ArrowLine(0,48)(17,38)
\DashArrowLine(17,38)(34,48){3}
\Text(1,17)[b]{$\scriptstyle H$}
\Text(33,17)[b]{$\scriptstyle \widetilde H$}
\Text(1,44)[t]{$\scriptstyle \ell$}
\Text(33,44)[t]{$\scriptstyle \widetilde \ell$}
\Text(13,30)[r]{$\scriptstyle \widetilde B$}
\Vertex(17,23){1.5}
\Vertex(17,38){1.5}

\SetOffset(142,130)

\Text(50,30)[]{$\scriptstyle \times$}
\Text(59,30)[]{$\Big($}
\DashArrowLine(64,20)(81,30){3}
\ArrowLine(64,40)(81,30)
\Line(81,30)(96,30)
\DashArrowLine(96,30)(113,40){3}
\ArrowLine(96,30)(113,20)
\Vertex(81,30){1.5}
\Vertex(96,30){1.5}
\Text(118,30)[]{$\Big)$}
\Text(120,40)[]{$\scriptstyle *$}
\Text(104,38)[b]{$\scriptstyle \widetilde \ell$}
\Text(104,22)[t]{$\scriptstyle \widetilde H$}
\Text(73,39)[b]{$\scriptstyle \ell$}
\Text(73,21)[t]{$\scriptstyle H$}
\Text(88,33)[b]{$\scriptstyle N$}


\SetOffset(0,65)
\Line(15,60)(245,60)

\Text(270,30)[l]{$\scriptstyle \textnormal{(A)}^*$}
\SetOffset(65,65)

\DashArrowLine(30,30)(0,30){3}
\ArrowLine(30,30)(56,15)
\Line(56,15)(82,30)
\Photon(56,15)(82,30){4}{3}
\Line(30,30)(56,45)
\ArrowLine(82,30)(56,45)
\DashArrowLine(112,30)(82,30){3}
\DashArrowLine(56,15)(56,45){3}
\Vertex(30,30){1.5}
\Vertex(56,15){1.5}
\Vertex(56,45){1.5}
\Vertex(82,30){1.5}
\Text(15,33)[b]{{$\scriptstyle \widetilde\ell$}}
\Text(97,33)[b]{{$\scriptstyle \widetilde\ell$}}
\Text(69,41)[b]{{$\scriptstyle \ell$}}
\Text(43,41)[b]{{$\scriptstyle N$}}
\Text(69,18)[t]{{$\scriptstyle \widetilde B$}}
\Text(43,18)[t]{{$\scriptstyle \widetilde H$}}
\Text(54,36)[r]{{$\scriptstyle H$}}

\SetColor{Red}
\Line(34,20)(34,40)
\Line(35,20)(35,40)
\SetColor{Orange}
\Line(36,25)(82,38)
\Line(36,26)(82,39)
\SetColor{Black}

\SetOffset(0,0)

\Text(120,60)[]{$\Longrightarrow$}

\Text(270,30)[l]{$\scriptstyle \textnormal{(B)}^*$}

\SetOffset(2,0)

\Line(0,30)(15,30)
\DashArrowLine(15,30)(41,15){3}
\ArrowLine(15,30)(41,45)
\Line(41,15)(41,45)
\Photon(41,15)(41,45){4}{3}
\DashArrowLine(41,45)(56,45){3}
\ArrowLine(41,15)(56,15)
\Vertex(15,30){1.5}
\Vertex(41,15){1.5}
\Vertex(41,45){1.5}
\Text(7,33)[b]{{$\scriptstyle N$}}
\Text(28,41)[b]{{$\scriptstyle\ell$}}
\Text(49,12)[t]{{$\scriptstyle \widetilde H$}}
\Text(49,47)[b]{{$\scriptstyle \widetilde \ell$}}
\Text(28,18)[t]{{$\scriptstyle H$}}
\Text(37,30)[r]{{$\scriptstyle \widetilde B$}}

\SetOffset(0,0)

\Text(72,30)[]{{$\scriptstyle \times$}}
\Text(0,30)[]{$\Bigg($}
\Text(64,30)[]{$\Bigg)$}
\Text(67,45)[]{$\scriptstyle *$}

\SetOffset(4,0)

\Line(74,30)(94,30)
\DashArrowLine(94,30)(111,40){3}
\ArrowLine(94,30)(111,20)
\Vertex(94,30){1.5}
\Text(84,33)[b]{{$\scriptstyle N$}}
\Text(103,40)[b]{{$\scriptstyle \widetilde\ell$}}
\Text(103,20)[t]{{$\scriptstyle \widetilde H$}}

\SetOffset(0,0)

\Text(127,30)[]{$\scriptstyle +$}

\SetOffset(147,0)

\DashArrowLine(0,13)(17,23){3}
\ArrowLine(17,23)(34,13)
\Line(17,23)(17,38)
\Photon(17,23)(17,38){3}{2.5}
\ArrowLine(0,48)(17,38)
\DashArrowLine(17,38)(34,48){3}
\Text(1,17)[b]{$\scriptstyle H$}
\Text(33,17)[b]{$\scriptstyle \widetilde H$}
\Text(1,44)[t]{$\scriptstyle \ell$}
\Text(33,44)[t]{$\scriptstyle \widetilde \ell$}
\Text(13,30)[r]{$\scriptstyle \widetilde B$}
\Vertex(17,23){1.5}
\Vertex(17,38){1.5}

\Text(-8,30)[]{$\Bigg($}
\Text(42,30)[]{$\Bigg)$}
\Text(45,45)[]{$\scriptstyle *$}

\Text(54,30)[]{$\scriptstyle \times$}

\SetOffset(144,0)

\DashArrowLine(64,20)(81,30){3}
\ArrowLine(64,40)(81,30)
\Line(81,30)(96,30)
\DashArrowLine(96,30)(113,40){3}
\ArrowLine(96,30)(113,20)
\Vertex(81,30){1.5}
\Vertex(96,30){1.5}
\Text(104,38)[b]{$\scriptstyle\widetilde \ell$}
\Text(104,22)[t]{$\scriptstyle \widetilde H$}
\Text(73,39)[b]{$\scriptstyle  \ell$}
\Text(73,21)[t]{$\scriptstyle H$}
\Text(88,33)[b]{$\scriptstyle N$}

\end{picture}
}
\end{center}
\caption{\label{fig:svertex:cuts} Relation of the slepton CTP self-energies
$\textnormal{(A)}$, $\textnormal{(A)}^*$
to the CPV interference of amplitudes
$\textnormal{(B)}$, $\textnormal{(B)}^*$
in conventional approaches,
on the example of the contribution from the off-shell $\widetilde B$,
where the star indicates that the diagrams represent terms that
are related by complex conjugation.
The CPV contributions to the interference terms occur when
$\ell$ and $H$ are on-shell. The RIS contribution
arises from an on-shell $N$ in the $2\leftrightarrow 2$ scattering amplitude. In the interference terms, we have suppressed the
notation of the
integration over the phase space of the external particles other
than $\widetilde \ell$.}
\end{figure}
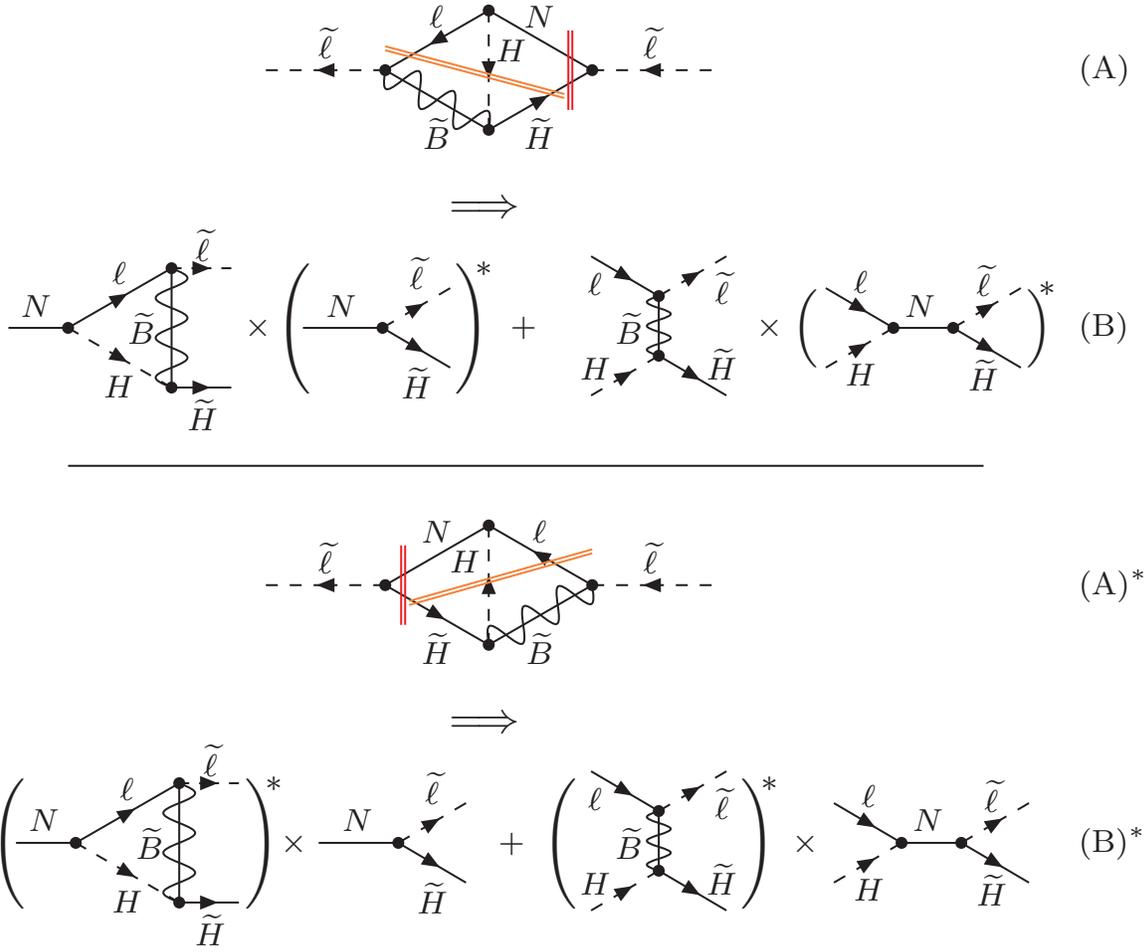

It is instructive to relate this cancellation to the conventional asymmetry computation and to identify the effects omitted in Ref.~\cite{Fong:2009iu}. To that end, we show in Figs.~\ref{fig:vertex:cuts}$\textnormal{(A)}$,~\ref{fig:vertex:cuts}$\textnormal{(A)}^*$ the cuts in the lepton self energy for an on-shell $N$, $H_1$, and $\ell$ and in Figs.~\ref{fig:vertex:cuts}$\textnormal{(B)}$,~\ref{fig:vertex:cuts}$\textnormal{(B)}^*$ the corresponding interfering one-loop and tree-level amplitudes that enter a conventional asymmetry computation. In Fig.~\ref{fig:svertex:cuts} we show the corresponding cuts and interfering amplitudes for the slepton self-energies for an on-shell $N$ ${\tilde\ell}$, and ${\tilde H}$. 
 The red cuts (color online) correspond to the interfering tree-level and vertex correction amplitudes for the $N$ decay, while the orange cuts (color online) encompass both the off-shell bino-cuts in the vertex graphs computed above as well as the CPV remainder after the RIS subtraction is implemented for the $2 \leftrightarrow 2$ processes.
In the procedure followed in Ref.~\cite{Fong:2009iu}, one would include statistical factors for the particles $\ell$ and $H$ in Figs.~\ref{fig:vertex:cuts}$\textnormal{(B)}$,~\ref{fig:vertex:cuts}$\textnormal{(B)}^*$ [sparticles $\widetilde \ell$ and $\widetilde H$ in Figs.~\ref{fig:svertex:cuts}$\textnormal{(B)}$,~\ref{fig:svertex:cuts}$\textnormal{(B)}^*$]but omit those for the sparticles $\widetilde \ell$ and $\widetilde H$ [particles $\ell$ and $H$ in Figs.~\ref{fig:svertex:cuts}$\textnormal{(B)}$,~\ref{fig:svertex:cuts}$\textnormal{(B)}^*$](no matter whether these appear as external states in the scattering amplitudes or as internal lines in the vertex diagrams). The result would then be a non-vanishing lepton number asymmetry at finite temperature. In contrast, the CTP computation consistently
includes the statistical factors for the internal lines as well, leading to the vanishing result of Eq.~(\ref{eq:cancel1}). 

This cancellation may also be regarded as a consequence of the fact that
the $d^3 k$ integration in Eq.~(\ref{eq:cancel1})
corresponds to closing the (s)lepton lines in
Fig.~\ref{fig:selfergs}, what leads to identical Feynman diagrams.
Therefore, no matter which mass relation -- $m_N>m_{\widetilde \ell}+m_{H1}$ or $m_N+m_{H1}<m_{\widetilde \ell}$ -- holds, the asymmetry produced within the
leptons will always cancel the asymmetry produced within the sleptons. The fact
that this also holds at finite temperature is in contrast to what is argued in
Ref.~\cite{Fong:2009iu}. The reason for the discrepancy is that here,
through the use of the CTP approach, no distinction
between internal propagators and external particles is made, and quantum  statistical 
corrections to both are applied. Therefore the
the cancellation of lepton and slepton asymmetries present in the
vacuum generalizes to finite temperature backgrounds as well.

In Appendix~\ref{appendix:coll:kinematics}, we show how
${\cal C}^{{\rm v}\widetilde B}_{\widetilde \ell}$ can be further
evaluated to take a form that is familiar from the collision
term in Boltzmann equations. Such a calculation is however not
necessary in order to demonstrate the cancellations, that are a main
topic of this work.

\subsection{Cuts with off-shell $H_1$ and off-shell $\widetilde H$}

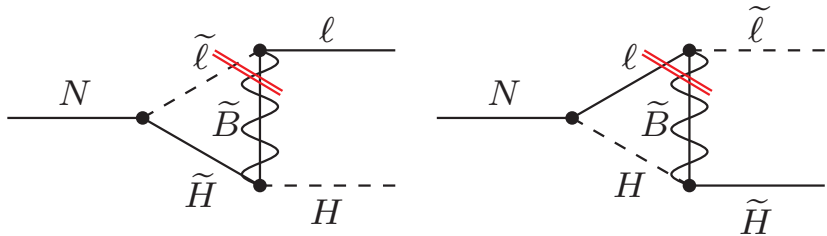
\begin{figure}[ht!]
\begin{center}
\scalebox{1.7}{
\begin{picture}(86,60)(0,0)

\Line(0,30)(30,30)
\Line(30,30)(56,15)
\DashLine(30,30)(56,45){3}
\Line(56,15)(56,45)
\Photon(56,15)(56,45){4}{3}
\Line(56,45)(86,45)
\DashLine(56,15)(86,15){3}
\Vertex(30,30){1.5}
\Vertex(56,15){1.5}
\Vertex(56,45){1.5}
\Text(15,33)[b]{{$\scriptstyle N$}}
\Text(43,41)[b]{{$\scriptstyle \widetilde\ell$}}
\Text(71,12)[t]{{$\scriptstyle H$}}
\Text(71,47)[b]{{$\scriptstyle \ell$}}
\Text(43,18)[t]{{$\scriptstyle \widetilde H$}}
\Text(52,30)[r]{{$\scriptstyle \widetilde B$}}

\SetColor{Red}
\Line(46,45)(61,36)
\Line(45.5,44.1)(60.5,35.1)

\end{picture}
}
\scalebox{1.7}{
\begin{picture}(86,60)(0,0)

\Line(0,30)(30,30)
\DashLine(30,30)(56,15){3}
\Line(30,30)(56,45)
\Line(56,15)(56,45)
\Photon(56,15)(56,45){4}{3}
\DashLine(56,45)(86,45){3}
\Line(56,15)(86,15)
\Vertex(30,30){1.5}
\Vertex(56,15){1.5}
\Vertex(56,45){1.5}
\Text(15,33)[b]{{$\scriptstyle N$}}
\Text(43,41)[b]{{$\scriptstyle \ell$}}
\Text(71,12)[t]{{$\scriptstyle \widetilde H$}}
\Text(71,47)[b]{{$\scriptstyle \widetilde\ell$}}
\Text(43,18)[t]{{$\scriptstyle H$}}
\Text(52,30)[r]{{$\scriptstyle \widetilde B$}}

\SetColor{Red}
\Line(46,45)(61,36)
\Line(45.5,44.1)(60.5,35.1)

\end{picture}
}
\end{center}
\caption{\label{fig:offhiggsinohiggs}Cuts in vacuum diagrams with off-shell Higgsino and Higgs boson.}
\end{figure}

It is also necessary to check whether the cancellation holds for the other
possible cuts.
We therefore consider contributions arising from terms where the
Higgs-boson $H_1$ is off-shell. From Fig.~\ref{fig:selfergs}
and Fig.~\ref{fig:offhiggsinohiggs}, we see that
the cuts where $\widetilde H$ and those where
$H_1$ are off-shell are qualitatively
similar, in the sense that they occur at a vertex with an
external particle that carries lepton numbers. Indeed, the expressions
for ${\cal C}^{{\rm v}\widetilde H1}_{\ell}$ and
${\cal C}^{{\rm v}\widetilde H1}_{\widetilde \ell}$ can be simply
inferred from the ones for
${\cal C}^{{\rm v}H1}_{\ell}$ and
${\cal C}^{{\rm v}H1}_{\widetilde \ell}$
that are derived in this Section by replacing the
quantum statistical factors. Consequently,  we do not present
the calculation for off-shell $\widetilde H$ here.

From the slepton self-energy~(\ref{Pi:ltilde}), we obtain
\begin{align}
\label{C:ltilde:phi}
{\cal C}_{\widetilde\ell}^{{\rm v}H1}
=&Y^2\frac{g_1^2}{4}{\rm tr}\int\frac{d k^0}{2\pi}\frac{d^4 p}{(2\pi)^4}\frac{d^4 q}{(2\pi)^4}
{\rm i}\Delta^T_{H1}(p+k+q)
\Big\{
\\\notag
&
\left[
P_{\rm L}{\rm i}S_{\widetilde B}^<(-p-k)
P_{\rm L}{\rm i}S_\ell^>(-p)
-P_{\rm L}{\rm i}S_{\widetilde B}^>(-p-k)
P_{\rm L}{\rm i}S_\ell^<(-p)
\right]
\\\notag
&
\times
\left[
P_{\rm R}{\rm i}S_N^>(q+k)
P_{\rm L}{\rm i}S_{\widetilde H}^<(q)
{\rm i}\Delta_{\widetilde \ell}^<(k)
-
P_{\rm R}{\rm i}S_N^<(q+k)
P_{\rm L}{\rm i}S_{\widetilde H}^>(q)
{\rm i}\Delta_{\widetilde \ell}^>(k)
\right]
\sin\alpha\cos\alpha {\rm e}^{-{\rm i}\phi_\mu}
\\\notag
&
-
\left[
P_{\rm L}{\rm i}S_\ell^>(-p)
P_{\rm L}{\rm i}S_{\widetilde B}^<(-p-k)
-P_{\rm L}{\rm i}S_\ell^<(-p)
P_{\rm L}{\rm i}S_{\widetilde B}^>(-p-k)
\right]
\\\notag
&
\times
\left[
P_{\rm R}{\rm i}S_{\widetilde H}^<(q)
P_{\rm R}{\rm i}S_N^>(q+k)
{\rm i}\Delta_{\widetilde \ell}^<(k)
-
P_{\rm R}{\rm i}S_{\widetilde H}^>(q)
P_{\rm R}{\rm i}S_N^<(q+k)
{\rm i}\Delta_{\widetilde \ell}^>(k)
\right]
\sin\alpha\cos\alpha {\rm e}^{{\rm i}\phi_\mu}
\\\notag
&+
\left[
P_{\rm L}{\rm i}S^<_{\widetilde B}(-p-k)
P_{\rm L}{\rm i}S^>_\ell(-p)
{\rm i}\Delta_{\widetilde\ell}^<(k)
-P_{\rm L}{\rm i}S^>_{\widetilde B}(-p-k)
P_{\rm L}{\rm i}S^<_\ell(-p)
{\rm i}\Delta_{\widetilde\ell}^>(k)
\right]
\\\notag
&
\times
\left[
P_{\rm R}{\rm i}S_{N}^<(q+k)
P_{\rm L}{\rm i}S^>_{\widetilde H}(q)
-P_{\rm R}{\rm i}S_N^>(q+k)
P_{\rm L}{\rm i}S^<_{\widetilde H}(q)
\right]
\sin\alpha\cos\alpha {\rm e}^{-{\rm i}\phi_\mu}
\\\notag
&-
\left[
P_{\rm L}{\rm i}S_\ell^>(-p)
P_{\rm L}{\rm i}S_{\widetilde B}^<(-p-k)
{\rm i}\Delta_{\widetilde\ell}^<(k)
-P_{\rm L}{\rm i}S_\ell^<(-p)
P_{\rm L}{\rm i}S_{\widetilde B}^>(-p-k)
{\rm i}\Delta_{\widetilde\ell}^>(k)
\right]
\\\notag
&
\times
\left[
P_{\rm R}{\rm i}S_{\widetilde H}^>(q)
P_{\rm R}{\rm i}S_N^<(q+k)
-
P_{\rm R}{\rm i}S_{\widetilde H}^<(q)
P_{\rm R}{\rm i}S_N^>(q+k)
\right]
\sin\alpha\cos\alpha {\rm e}^{{\rm i}\phi_\mu}
\Big\}
\,.
\end{align}
We note that the last two terms are vanishing due to KMS relations
applied to $S_\ell$, $S_{\widetilde B}$ and $\Delta_{\widetilde \ell}$.
Applying KMS relations to $S_{\widetilde B}$, $S_\ell$ and $S_N$ would
also render the first two terms vanishing. Since we consider however
situations when $N$ is out-of-equilibrium, we do not apply KMS
in this case, such that these terms remain as contributions to
the asymmetry in sleptons $\widetilde \ell$.


The lepton collision term with the off-shell $H_1$ takes a form
that is diagrammatically different from the slepton collision term~(\ref{C:ltilde:phi}), in the sense that here $\Delta_{H1}$ connects to
an external vertex of the self-energy $\slashed\Sigma^{\rm v}_\ell$, while
for $\Pi^{\rm v}_{\widetilde \ell}$, it connects two internal
vertices.
We obtain
\begin{align}
\label{C:ltilde:H1}
\int\frac{d^3 k}{(2\pi)^3}{\cal C}_\ell^{{\rm v}H1}=&
-Y^2\frac{g_1^2}{2}\cos\alpha\sin\alpha\,
{\rm tr}
\int\frac{d^4k}{(2\pi)^4}
\frac{d^4p}{(2\pi)^4}
\frac{d^4 q}{(2\pi)^4}
{\rm i}\Delta^T_{H1}(q)
\Big\{
\\\notag
&
\big[
{\rm i}S_\ell^<(k)
P_{\rm R}{\rm i}S_{\widetilde B}^>(p+k)
P_{\rm R}{\rm i}S_{\widetilde H}^<(p+k+q)
P_{\rm R}{\rm i}S_N^>(q+k)
\\\notag
&
-
{\rm i}S_\ell^>(k)
P_{\rm R}{\rm i}S_{\widetilde B}^<(p+k)
P_{\rm R}{\rm i}S_{\widetilde H}^>(p+k+q)
P_{\rm R}{\rm i}S_N^<(q+k)
\big]
\\\notag
&\times
\left(
{\rm i}\Delta^>_{\widetilde \ell}(-p)-{\rm i}\Delta^{\bar T}_{\widetilde \ell}(-p)
\right){\rm e}^{{\rm i}\phi_\mu}
\\\notag
&
-
\big[
P_{\rm L}{\rm i}S_{\widetilde B}^>(p+k)
P_{\rm L}{\rm i}S_\ell^<(k)
{\rm i}S_N^>(q+k)
P_{\rm L}{\rm i}S_{\widetilde H}^<(p+k+q)
\\\notag
&-
P_{\rm L}{\rm i}S_{\widetilde B}^<(p+k)
P_{\rm L}{\rm i}S_\ell^>(k)
{\rm i}S_N^<(q+k)
P_{\rm L}{\rm i}S_{\widetilde H}^>(p+k+q)
\big]
\\\notag
&\times
\left(
-{\rm i}\Delta^>_{\widetilde \ell}(-p)+{\rm i}\Delta^{T}_{\widetilde \ell}(-p)
\right){\rm e}^{-{\rm i}\phi_\mu}
\Big\}\,.
\end{align}
Note that when the neutrino $N$ is in equilibrium, the integrand vanishes, as can
be easily verified by applying KMS relations to the terms in square brackets. Next,
we notice that the terms in square brackets are odd in the momentum variables (this is a consequence of
$\ell$, $\widetilde B$, $\widetilde H$ being in equilibrium and $N$ being its own anti-particle),
while the terms in round brackets are the retarded/advanced propagators ${\rm i}\Delta^R={\rm i}\Delta^>-{\rm i}\Delta^{\bar T}=-{\rm i}\Delta^<+{\rm i}\Delta^{T}$,
${\rm i}\Delta^A={\rm i}\Delta^<-{\rm i}\Delta^{\bar T}=-{\rm i}\Delta^>+{\rm i}\Delta^{T}$]. The components ${\rm i}\Delta^T$ and ${\rm i}\Delta^{\bar T}$
are even in the momentum variable and therefore yield a vanishing contribution
to the integral.
Making use of this observation, we can reexpress
\begin{align}
\int\frac{d^3 k}{(2\pi)^3}{\cal C}_\ell^{{\rm v}H1}=&
-Y^2\frac{g_1^2}{2}\cos\alpha\sin\alpha\,
{\rm tr}
\int\frac{d^4k}{(2\pi)^4}
\frac{d^4p}{(2\pi)^4}
\frac{d^4 q}{(2\pi)^4}
{\rm i}\Delta^T_{H1}(q)
\Big\{
\\\notag
&
{\rm i}S_\ell^>(k)
P_{\rm R}{\rm i}S_{\widetilde B}^<(p+k)
\big[
P_{\rm R}{\rm i}S_{\widetilde H}^<(p+k+q)
P_{\rm R}{\rm i}S_N^>(q+k)
{\rm i}\Delta_{\widetilde \ell}^<(-p)
\\\notag
&
-
P_{\rm R}{\rm i}S_{\widetilde H}^>(p+k+q)
P_{\rm R}{\rm i}S_N^<(q+k)
{\rm i}\Delta_{\widetilde \ell}^>(-p)
\big]
{\rm e}^{{\rm i}\phi_\mu}
\\\notag
&
-
P_{\rm L}{\rm i}S_{\widetilde B}^<(p+k)
{\rm i}S_\ell^>(k)
\big[
P_{\rm R}{\rm i}S_N^>(q+k)
P_{\rm L}{\rm i}S_{\widetilde H}^<(p+k+q)
{\rm i}\Delta_{\widetilde \ell}^<(-p)
\\\notag
&
-
P_{\rm R}{\rm i}S_N^<(q+k)
P_{\rm L}{\rm i}S_{\widetilde H}^>(p+k+q)
{\rm i}\Delta_{\widetilde \ell}^>(-p)
\big]
{\rm e}^{-{\rm i}\phi_\mu}
\Big\}
\\\notag
\underset{\text{(KMS)}}{=}&
-Y^2\frac{g_1^2}{2}\cos\alpha\sin\alpha\,
{\rm tr}
\int\frac{d^4k}{(2\pi)^4}
\frac{d^4p}{(2\pi)^4}
\frac{d^4 q}{(2\pi)^4}
{\rm i}\Delta^T_{H1}(q)
\Big\{
\\\notag
&
\big[
{\rm i}S_\ell^<(k)
P_{\rm R}{\rm i}S_{\widetilde B}^>(p+k)
P_{\rm R}{\rm i}S_{\widetilde H}^<(p+k+q)
P_{\rm R}{\rm i}S_N^>(q+k)
\\\notag
&
-
{\rm i}S_\ell^>(k)
P_{\rm R}{\rm i}S_{\widetilde B}^<(p+k)
P_{\rm R}{\rm i}S_{\widetilde H}^>(p+k+q)
P_{\rm R}{\rm i}S_N^<(q+k)
\big]{\rm i}\Delta_{\widetilde \ell}^>{\rm e}^{{\rm i}\phi_\mu}
\\\notag
&
-
\big[
P_{\rm L}{\rm i}S_{\widetilde B}^>(p+k)
{\rm i}S_\ell^<(k)
P_{\rm R}{\rm i}S_N^>(q+k)
P_{\rm L}{\rm i}S_{\widetilde H}^<(p+k+q)
\\\notag
&
-
P_{\rm L}{\rm i}S_{\widetilde B}^<(p+k)
{\rm i}S_\ell^>(k)
P_{\rm R}{\rm i}S_N^<(q+k)
P_{\rm L}{\rm i}S_{\widetilde H}^>(p+k+q)
\big]{\rm i}\Delta_{\widetilde \ell}^>{\rm e}^{-{\rm i}\phi_\mu}
\Big\}
\,,
\end{align}
where we have applied the KMS relation to $S_{\ell}$, $S_{\widetilde B}$
and $\Delta_{\widetilde \ell}$.
Again, by appealing to the symmetry property in the momentum variables
(terms are even under the simultaneous reversal of the momenta and the
replacement
$<\leftrightarrow>$), we may replace
\begin{align}
&\left(
S_\ell^<S_{\widetilde B}^>S_{\widetilde H}^< S_N^>
-
S_\ell^>S_{\widetilde B}^<S_{\widetilde H}^> S_N^<
\right)\Delta_{\widetilde \ell}^>
\rightarrow
\\\notag
&\frac12\left(
S_\ell^<S_{\widetilde B}^>S_{\widetilde H}^< S_N^>
-
S_\ell^>S_{\widetilde B}^<S_{\widetilde H}^> S_N^<
\right)
\left(\Delta_{\widetilde \ell}^>-\Delta_{\widetilde \ell}^<\right)
\\\notag
\underset{\text{(KMS)}}{=}&
\frac12
\left(S_\ell^>S_{\widetilde B}^<-S_\ell^<S_{\widetilde B}^>\right)
\times
\left(
S_{\widetilde H}^<S_N^>\Delta_{\widetilde \ell}^<
-S_{\widetilde H}^>S_N^<\Delta_{\widetilde \ell}^>
\right)
\,,
\end{align}
which then leads us to the observation
\begin{align}
\int\frac{d^3k}{(2\pi)^3}\left[
{\cal C}_{\widetilde \ell}^{{\rm v}H1}(\mathbf k)+{\cal C}_\ell^{{\rm v}H1}(\mathbf k)
\right]=0
\,.
\end{align}
Therefore, provided superequilibrium (equilibrium between
superpartners) holds, there will be no residual lepton asymmetry.

A useful check is to suppose that instead of a Majorana mass, $N$ would have a Dirac mass, which can 
be achieved by adding another chiral sterile neutrino degree of freedom. Then the collision
term for creation of an $N$-charge must vanish, which can also
be verified explicitly.

\subsection{Cuts with off-shell $\widetilde \ell$}

\label{section:ltildeoff}

Finally,  cuts can be applied to the vertex graphs in such a way that they do
not go through a line that carries lepton number. We consider
here the situation when the slepton $\widetilde \ell$ is off-shell.
The analogous result for an off-shell lepton $\ell$ can be easily
inferred from what is presented here.

\begin{figure}[ht!]
\begin{center}
\scalebox{1.7}{
\begin{picture}(86,60)(0,0)

\Line(0,30)(30,30)
\Line(30,30)(56,15)
\DashLine(30,30)(56,45){3}
\Line(56,15)(56,45)
\Photon(56,15)(56,45){4}{3}
\Line(56,45)(86,45)
\DashLine(56,15)(86,15){3}
\Vertex(30,30){1.5}
\Vertex(56,15){1.5}
\Vertex(56,45){1.5}
\Text(15,33)[b]{{$\scriptstyle N$}}
\Text(43,41)[b]{{$\scriptstyle \widetilde\ell$}}
\Text(71,12)[t]{{$\scriptstyle H$}}
\Text(71,47)[b]{{$\scriptstyle \ell$}}
\Text(43,18)[t]{{$\scriptstyle \widetilde H$}}
\Text(52,30)[r]{{$\scriptstyle \widetilde B$}}

\SetColor{Red}
\Line(46,15)(61,24)
\Line(45.5,15.9)(60.5,24.9)

\end{picture}
}
\scalebox{1.7}{
\begin{picture}(86,60)(0,0)

\Line(0,30)(30,30)
\DashLine(30,30)(56,15){3}
\Line(30,30)(56,45)
\Line(56,15)(56,45)
\Photon(56,15)(56,45){4}{3}
\DashLine(56,45)(86,45){3}
\Line(56,15)(86,15)
\Vertex(30,30){1.5}
\Vertex(56,15){1.5}
\Vertex(56,45){1.5}
\Text(15,33)[b]{{$\scriptstyle N$}}
\Text(43,41)[b]{{$\scriptstyle \ell$}}
\Text(71,12)[t]{{$\scriptstyle \widetilde H$}}
\Text(71,47)[b]{{$\scriptstyle \widetilde\ell$}}
\Text(43,18)[t]{{$\scriptstyle H$}}
\Text(52,30)[r]{{$\scriptstyle \widetilde B$}}

\SetColor{Red}
\Line(46,15)(61,24)
\Line(45.5,15.9)(60.5,24.9)

\end{picture}
}
\end{center}
\caption{\label{fig:offelltildeell}Cuts in vacuum diagrams with off-shell slepton and lepton.}
\end{figure}
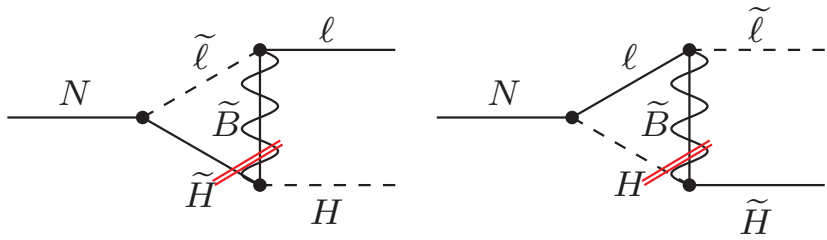

For the CPV vertex contribution to the lepton
collision term, we obtain
\begin{align}
\int\frac{d^3k}{(2\pi)^3}{\cal C}_\ell^{{\rm v}\widetilde \ell}(\mathbf k)
=&
Y^2\frac{g_1^2}{2}\cos\alpha\sin\alpha
\int\frac{d^4k}{(2\pi)^4}\frac{d^4p}{(2\pi)^4}\frac{d^4q}{(2\pi)^4}
{\rm i}\Delta_{\widetilde \ell}^T(-p)
\Big\{
\\\notag
&
\big[
P_{\rm R}{\rm i}S_{\widetilde B}^>(p+k)
P_{\rm R}{\rm i}S_{\widetilde H}^<(p+k+q)
P_{\rm R}{\rm i}S_N^>(q+k)
P_{\rm L}{\rm i}S_\ell^<(k)
\\\notag
&
-P_{\rm R}{\rm i}S_{\widetilde B}^<(p+k)
P_{\rm R}{\rm i}S_{\widetilde H}^>(p+k+q)
P_{\rm R}{\rm i}S_N^<(q+k)
P_{\rm L}{\rm i}S_\ell^>(k)
\big]
{\rm i}\Delta_{H1}^H(q){\rm e}^{{\rm i}\phi_\mu}
\\\notag
&
+
\big[
P_{\rm L}{\rm i}S_{\ell}^<(k)
P_{\rm R}{\rm i}S_{N}^>(q+k)
P_{\rm L}{\rm i}S_{\widetilde H}^<(p+k+q)
P_{\rm L}{\rm i}S_{\widetilde B}^>(p+k)
\\\notag
&
-P_{\rm L}{\rm i}S_{\ell}^>(k)
P_{\rm R}{\rm i}S_{N}^<(q+k)
P_{\rm L}{\rm i}S_{\widetilde H}^>(p+k+q)
P_{\rm L}{\rm i}S_{\widetilde B}^<(p+k)
\big]
{\rm i}\Delta_{H1}^H(q){\rm e}^{-{\rm i}\phi_\mu}
\Big\}\,,
\end{align}
where ${\rm i}\Delta^H=\frac12({\rm i}\Delta^A+{\rm i}\Delta^R)
=\frac12({\rm i}\Delta^T-{\rm i}\Delta^{\bar T})$.
By KMS again, the integrand vanishes when $N$ is in equilibrium. Similar to the
vanishing of ${\cal C}_{\widetilde \ell}^{{\rm v}H1}+{\cal C}_\ell^{{\rm v}H1}$, the integrated collision term also vanishes
when $N$ is not in equilibrium. This is because
${\rm i}\Delta^H(k)={\rm i}\Delta^H(-k)$ whereas the
terms in the square bracket are odd under momentum reversal.

\section{Cancellation of Contributions from Wavefunction Diagrams}

\label{section:wv}

\begin{figure}[ht!]
\begin{center}
\scalebox{1.7}{
\begin{picture}(270,135)(0,0)

\SetOffset(80,70)
\ArrowLine(30,40)(0,40)
\CArc(55,40)(25,0,180)
\ArrowLine(110,40)(80,40)
\DashArrowArc(55,40)(25,180,238){3}
\DashArrowArc(55,40)(25,302,360){3}
\ArrowArc(55,19)(13,180,360)
\CArc(55,19)(13,0,180)
\PhotonArc(55,19)(13,0,180){-4}{3.5}
\Vertex(30,40){1.5}
\Vertex(42,19){1.5}
\Vertex(68,19){1.5}
\Vertex(80,40){1.5}
\Text(15,43)[b]{$\scriptstyle \ell$}
\Text(95,43)[b]{$\scriptstyle \ell$}
\Text(55,9)[b]{$\scriptstyle \widetilde H$}
\Text(55,38)[b]{$\scriptstyle \widetilde B$}
\Text(55,62)[t]{$\scriptstyle N$}
\Text(30,25)[t]{$\scriptstyle H_2$}
\Text(82,25)[t]{$\scriptstyle H_1$}

\Text(55,-6)[]{$\Longrightarrow$}

\SetOffset(0,0)

\Line(0,30)(20,30)
\ArrowLine(20,30)(53,54)
\DashArrowLine(20,30)(30,23){3}
\DashArrowLine(43,14)(53,6){3}
\Vertex(20,30){1.5}
\Vertex(31,22.5){1.5}
\Vertex(42,14.5){1.5}
\Text(10,33)[b]{{$\scriptstyle N$}}
\Text(32,43)[b]{{$\scriptstyle \ell$}}
\Text(23,24)[t]{{$\scriptstyle H_2$}}
\Text(46,8)[t]{{$\scriptstyle H_1$}}
\Text(31,11)[t]{$\scriptstyle \widetilde H$}
\Text(44,27)[b]{$\scriptstyle \widetilde B$}
\ArrowArc(37,19)(7,145,325)
\CArc(37,19)(7,325,145)
\PhotonArc(37,19)(7,325,145){2}{3.5}

\Text(58,30)[]{$\scriptstyle \times$}

\Text(65,30)[]{$\bigg($}

\Line(67,30)(87,30)
\ArrowLine(87,30)(104,40)
\DashArrowLine(87,30)(104,20){3}
\Vertex(87,30){1.5}
\Text(77,33)[b]{{$\scriptstyle N$}}
\Text(96,38)[b]{{$\scriptstyle \ell$}}
\Text(96,20)[t]{{$\scriptstyle H_1$}}

\Text(109,30)[]{$\bigg)$}
\Text(112,42)[]{$\scriptstyle *$}

\Text(120,30)[]{$\scriptstyle +$}

\SetOffset(128,0)

\Line(0,30)(20,30)
\ArrowLine(20,30)(46,45)
\DashArrowLine(20,30)(33,22){3}
\ArrowLine(33,22)(46,14)
\Line(33,22)(46,30)
\Photon(33,22)(46,30){2.5}{2.5}
\Text(10,33)[b]{$\scriptstyle N$}
\Text(31,41)[b]{$\scriptstyle \ell$}
\Text(25,23)[t]{$\scriptstyle H_2$}
\Text(46,30)[l]{$\scriptstyle \widetilde B$}
\Text(46,14)[l]{$\scriptstyle \widetilde H$}
\Vertex(20,30){1.5}
\Vertex(33,22){1.5}

\Text(60,30)[]{$\scriptstyle \times$}

\SetOffset(198,0)

\Text(-3,30)[]{$\Bigg($}

\Line(0,30)(20,30)
\ArrowLine(20,30)(46,45)
\DashArrowLine(20,30)(33,22){3}
\ArrowLine(33,22)(46,14)
\Line(33,22)(46,30)
\Photon(33,22)(46,30){2.5}{2.5}
\Text(10,33)[b]{$\scriptstyle N$}
\Text(31,41)[b]{$\scriptstyle \ell$}
\Text(25,23)[t]{$\scriptstyle H_1$}
\Text(46,30)[l]{$\scriptstyle \widetilde B$}
\Text(46,14)[l]{$\scriptstyle \widetilde H$}
\Vertex(20,30){1.5}
\Vertex(33,22){1.5}

\Text(59,30)[]{$\Bigg)$}
\Text(62,46)[]{$\scriptstyle *$}

\end{picture}
}
\end{center}
\caption{\label{fig:offhiggswv}Diagrams that lead to
potential contributions to the lepton asymmetry but that are
vanishing when $H$, $\widetilde H$ and $\widetilde B$ are in
equilibrium. In the interference terms, we have suppressed
the notation of the integrals over the phase space of the external
particles other than $\ell$. Notice that when $H_1$ corresponds
to a RIS, the $1\leftrightarrow 3$ amplitudes
can be cut along $H_1$ and fused along $\widetilde B$ and
$\widetilde H$, such that they cancel the interference
of the $1\leftrightarrow 2$ amplitudes, as it is explained in
Section~\ref{sec:RIS}.
}
\end{figure}
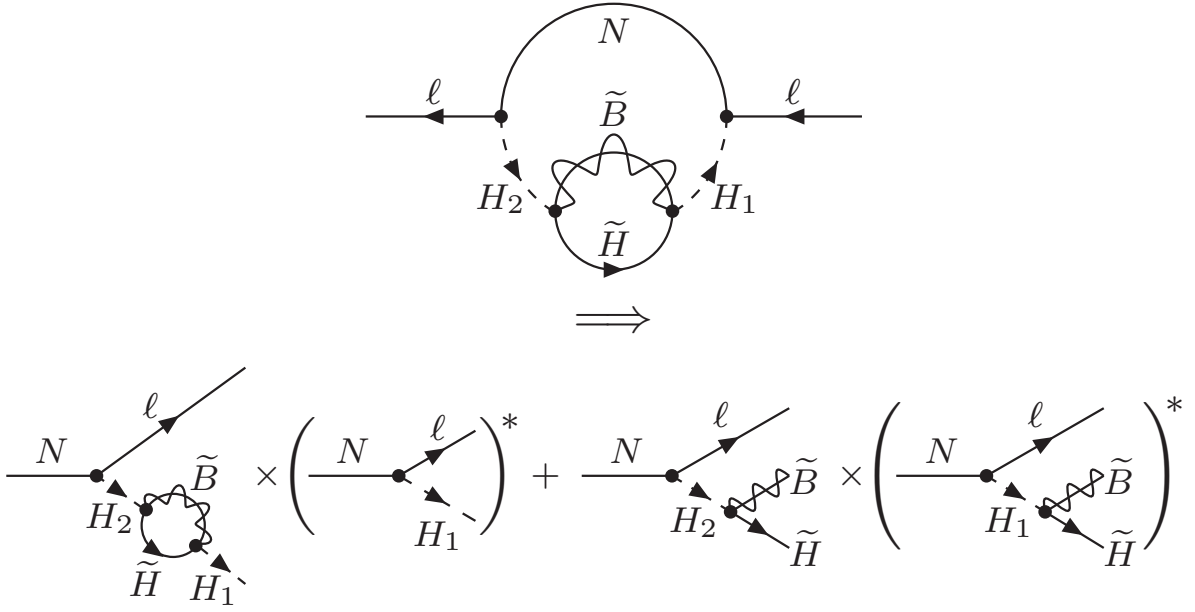

In the context of soft leptogenesis, another potentially significant contribution 
arises from CP-violation in the mixing of $H_1$ and $H_2$ that is
induced through loops with $\widetilde B$ and
$\widetilde H$. While the exclusive
vacuum decay rate of a singlet neutrino
$N$ to $H_1$ and $\ell$ receives CPV contributions
through the first diagram in Fig.~\ref{fig:offhiggswv},
using the CTP formalism, we can convince ourselves
that there is no contribution to leptogenesis, though.

For simplicity, let us assume that $m_{H2}\gg m_{H1}$
and $m_{H2}\gg T$, such that
$H_2$ is always off-shell to a good approximation. The more general
case can be easily inferred from the following results.
The CPV wave-function contribution
to the lepton collision term, that results from the lepton self-energy
diagram in Fig.~\ref{fig:offhiggswv}, is
then given by
\begin{align}
{\cal C}^{\rm w}_\ell(\mathbf k)
&=\frac{g_1^2}{2}Y^2\sin\alpha\cos\alpha\sin\phi_\mu
\int\frac{dk^0}{2\pi}\frac{d^4 p}{(2\pi)^4}\frac{d^4 q}{(2 \pi)^4}
\frac{1}{p^2-m_{H2}^2}
\\\notag
&\times
{\rm tr}\left[
P_{\rm R}{\rm i}S_N^<(p+k){\rm i}S_\ell^>(k)
-P_{\rm R}{\rm i}S_N^>(p+k){\rm i}S_\ell^<(k)
\right]
\\\notag
&\times
{\rm tr}\left[
{\rm i}\Delta_{H1}^<(p)
P_{\rm R}
{\rm i}S_{\widetilde B}^<(q)
P_{\rm R}
{\rm i}S_{\widetilde H}^>(p+q)
-
{\rm i}\Delta_{H1}^>(p)
P_{\rm R}
{\rm i}S_{\widetilde B}^>(q)
P_{\rm R}
{\rm i}S_{\widetilde H}^<(p+q)
\right]
\,.
\end{align}
In order to derive this expression, no use of the KMS relation has
been made. However, it is obvious that when applying KMS to
the propagators $\Delta_{H1}$, $S_{\widetilde B}$ and $S_{\widetilde H}$,
it immediately follows
that ${\cal C}^{\rm w}_\ell(\mathbf k)\equiv 0$. Essentially,
this cancellation is closely related to the fact that in vacuum,
the inclusive decay rate of $N$, which is obtained when considering
both $\ell + H_1$ also $\ell +\widetilde B+\widetilde H$
as final states leads to a vanishing contribution to the  lepton asymmetry.

\section{Relation to RIS Subtraction}
\label{sec:RIS}

The results of Sections~\ref{section:ltildeoff} and~\ref{section:wv}
indicate that the asymmetries in $\ell$ associated with off-shell  $\widetilde \ell$ vertex graphs and with
wavefunction diagrams readily vanish, even
without a cancellation with $\widetilde \ell$. When finite temperature
effects are neglected, the vanishing results can also be explained
using the standard method of Boltzmann equations supplemented
by the subtraction of RIS. We first review the salient features
of the subtraction procedure for standard leptogenesis and
then apply it to the process that is calculated in
Section~\ref{section:wv}.

In  leptogenesis calculations that rely on Boltzmann equations
in the limit of non-relativistic singlet neutrinos (i.e.
$m_N\gg T$), it is useful to
employ the averaged decay rate~\cite{Giudice:2003jh}
\begin{align}
\label{gamma:av}
\gamma^{\rm av}=
\Gamma_{N\to \ell H,\bar\ell H^*}
\frac
{
\int\frac{d^3 p}{(2\pi)^3}
\frac{m_N}{\sqrt{\mathbf p^2+m_N^2}}
{\rm e}^{-\sqrt{\mathbf p^2+m_N^2}/T}
}
{
\int\frac{d^3 p}{(2\pi)^3}
{\rm e}^{-\sqrt{\mathbf p^2+m_N^2}/T}
}
=
\frac{K_1(z)}{K_2(z)}
\Gamma_{N\to \ell H,\bar\ell H^*}
\,,
\end{align}
where the factor $m_N/E_N$ accounts for time dilation, 
$z=m_N/T$ and $\Gamma_{N\to \ell H,\bar\ell H^*}$ is the
total vacuum decay rate of a singlet neutrino $N$ into a
lepton $\ell$ and Higgs boson $H$ and their
anti-particles. Note that while this expression includes the relativistic time-dilation factors, but it is non-relativistic
in the sense that the quantum-statistical distributions are
replaced by classical Maxwell distributions.\footnote{For completeness, even though this is not essential for the
arguments presented in this Section, we note that
the relation to the conformal time $\eta$ used
for the collision rates throughout the remainder of
this paper is given by
$T=1/\eta$ for a scale factor in the radiation-dominated
Universe $a(\eta)=m_{\rm Pl} \eta \sqrt{45/(\pi^3 g_\star)}/2$.
The Hubble rate then is $H=(da(\eta)/d\eta)/a^2(\eta)$. In the
above expression for the collision terms, one should then
multiply the mass terms by $a(\eta)$ and replace the temperature
by the constant comoving temperature
$T_{\rm com}=(a_{\rm R}m_{\rm Pl}/2)^{1/2}(45/\pi^3 g_\star)^{1/4}$.
More details of this parametrization, that is useful in the
CTP approach, are given in Refs.~\cite{Beneke:2010wd,Garbrecht:2011aw}.}

We furthermore note the standard definition $Y_X=n_X/s$, where
$n_X$ is the number density of the particle $X$ and $s$ is the
entropy density. Furthermore, $Y_X^{\rm eq}$ denotes the value that
$Y_X$ takes in thermal equilibrium.
Using the non-relativistic, averaged decay rate~(\ref{gamma:av}), the
Boltzmann equations for leptogenesis can be expressed as
\begin{subequations}
\begin{align}
zHs\frac{dY_N}{dz}=&-\gamma^{\rm av}\frac{Y_N-Y_N^{\rm eq}}{Y_N^{\rm eq}}
\,,\\
\label{BE:ql}
zHs\frac{d(Y_{\ell}-Y_{\bar \ell})}{dz}=&
\frac{Y_N}{Y_N^{\rm eq}}\frac{1+\varepsilon}{2}\gamma^{\rm av}
-\frac{Y_N}{Y_N^{\rm eq}}\frac{1-\varepsilon}{2}\gamma^{\rm av}
+\frac{Y_{\bar\ell}}{Y_\ell^{\rm eq}}\frac{1+\varepsilon}{2}\gamma^{\rm av}
-\frac{Y_{\ell}}{Y_\ell^{\rm eq}}\frac{1-\varepsilon}{2}\gamma^{\rm av}
\\\notag
-&2\gamma_{\ell H\to \bar \ell H^*}
+2\gamma_{\bar \ell H^* \to \ell H}
+2\gamma^{\rm RIS}_{\ell H\to \bar \ell H^*}
-2\gamma^{\rm RIS}_{\bar \ell H^* \to \ell H}\,.
\end{align}
\end{subequations}
Here, $\varepsilon$ is the usual parameter that describes the
relative asymmetry in vacuum decays of $N$. The
relative signs in front of it
follow either from direct calculation or can easily be inferred
using general arguments of charge conjugation and the CPT-invariance
theorem. The factors of $1/2$ arise because the individual  decay rates into
$\ell H$ and ${\bar\ell} H^\ast$ final states are equal in the absence of CPV, and $\gamma^{\rm av}$ accounts for the averaged total decay rate
into both of these states.

The particular terms on the right hand side of
Eq.~(\ref{BE:ql}) are explained as follows: The first two
describe asymmetric decays of $N$ into $\ell$ and $\bar \ell$, while
the third and fourth term describe inverse decays. One might expect that
it is sufficient to consider these $1\leftrightarrow2$ processes in
order to describe leptogenesis at leading order.
However, the $2\leftrightarrow2$ $N$-mediated
scatterings between $\ell H$
and $\bar\ell H^*$ contribute to lepton number violation at
leading order when performing the phase space integral over
an on-shell intermediate singlet neutrino $N$ in the $s$-channel.
Moreover, for 
 $Y_\ell=Y_{\bar\ell}=Y_\ell^\mathrm{eq}$
the first four terms then add up to
$\varepsilon(Y_N/Y_N^{\rm eq}+1)\gamma^{\rm av}$, indicating that
an asymmetry would even be generated in equilibrium
 in conflict with the requirements of CPT invariance.
Therefore, the fifth and sixth
terms must be included, which correspond to $2\leftrightarrow 2$ scattering
processes mediated by and $s$-channel $N$. In other words, the Boltzmann
equations are completed by attaching external $\ell$ and $H$ to the unstable $N$ --
which also give rise to the relevant CPV contributions.
The $2\leftrightarrow 2$ rates
are CP-invariant, but they include regions of phase space
where there is an $s$-channel divergence of the propagator of $N$.
The rates for reactions via
these so-called real intermediate states (RIS) are already
accounted in the $1\leftrightarrow 2$ processes and must be
subtracted~\cite{Kolb:1979qa,Giudice:2003jh}
and can be written in the following suggestive way:
\begin{subequations}
\begin{align}
\gamma^{\rm RIS}_{\ell H\to \bar \ell H^*}
=&\frac{Y_\ell}{Y_\ell^{\rm eq}}\times\frac{1-\varepsilon}{2}\gamma^{\rm av}
\times\frac{1-\varepsilon}{2}
\approx\gamma^{\rm av}\frac{1-2\varepsilon}{4}\,,\\
\gamma^{\rm RIS}_{\bar\ell H^* \to \ell H}
=&\frac{Y_{\bar \ell}}{Y_\ell^{\rm eq}}\times\frac{1+\varepsilon}{2}\gamma^{\rm av}
\times\frac{1+\varepsilon}{2}
\approx\gamma^{\rm av}\frac{1+2\varepsilon}{4}\,.
\end{align}
\end{subequations}
The first factor involving $\varepsilon$ is the CPV
inverse decay rate of $N$, while the second factor is the
branching ratio of the CPV decays. Substitution
into Eq.~(\ref{BE:ql}) leads to the production rate
$\varepsilon(Y_N/Y_N^{\rm eq}-1)\gamma^{\rm av}$ for the lepton asymmetry,
which vanishes in equilibrium, as it should. Note that
the washout terms remain present within the unsubtracted, $CP$-conserving
$2\leftrightarrow 2$ rates, which are the
5th and 6th term on the right hand side of Eq.~(\ref{BE:ql}).

As emphasized above, a perhaps less heuristic and more controlled way to achieve this
result is using the CTP approach, i.e.  the same methods
employed in the remainder of this paper, and as it is
exercised in Ref.~\cite{Beneke:2010wd}
(see  also
Refs.~\cite{Garny:2009qn,Garny:2009rv} for purely scalar models). Nonetheless, it is 
instructive to identify the origin of the vanishing contributions of Sections~\ref{section:ltildeoff} and~\ref{section:wv}
in the context of the conventional  Boltzmann framework. In so doing, we emphasize that the analysis of Sections~\ref{section:ltildeoff} and~\ref{section:wv} are more general
because the argument based on RIS subtraction assumes a vacuum background,
which is an appropriate approximation only when $m_N\gg T$.
Moreover, we assume here that $m_N>m_\ell+m_{H1}$ and
$m_{H1}>m_{\widetilde H}+m_{\widetilde B}$, while the
results of Sections~\ref{section:ltildeoff} and~\ref{section:wv}
apply to more general kinematic situations as well.

With these comments in mind, we write down the production rate for the lepton asymmetry
\begin{align}
\label{BE:ql:softlepto}
zHs\frac{d(Y_{\ell}-Y_{\bar \ell})}{dz}=&
\frac{Y_N}{Y_N^{\rm eq}}\frac{1+\varepsilon}{2}\gamma^{\rm av}
-\frac{Y_N}{Y_N^{\rm eq}}\frac{1-\varepsilon}{2}\gamma^{\rm av}
+\frac{Y_{\bar\ell}}{Y_\ell^{\rm eq}}\frac{1+\varepsilon}{2}\gamma^{\rm av}
-\frac{Y_{\ell}}{Y_\ell^{\rm eq}}\frac{1-\varepsilon}{2}\gamma^{\rm av}
\\\notag
-&
\gamma_{\ell\widetilde H\widetilde B\to N}
+\gamma_{N\to\ell\widetilde H\widetilde B}
+\gamma_{\bar\ell\bar{\widetilde H}\widetilde B\to N}
-\gamma_{N\to\bar\ell\bar{\widetilde H}\widetilde B}
\\\notag
+&\gamma^{\rm RIS}_{\ell\widetilde H\widetilde B\to N}
-\gamma^{\rm RIS}_{N\to\ell\widetilde H\widetilde B}
-\gamma^{\rm RIS}_{\bar\ell\bar{\widetilde H}\widetilde B\to N}
+\gamma^{\rm RIS}_{N\to\bar\ell\bar{\widetilde H}\widetilde B}
\,.
\end{align}

For the wavefunction contributions (Section~\ref{section:wv}),
the $1\leftrightarrow 3$ processes are mediated by $H_1$ or $H_2$, where
$H_1$ may be on-shell. Interferences between $H_1$ and $H_2$ mediated
processes correspond to a single cut through the $N$, ${\widetilde B}$, and ${\widetilde H}$ lines in Fig.~\ref{fig:offhiggswv}.\footnote{
For the vertex contributions with off-shell $\widetilde \ell$
(Section~\ref{section:ltildeoff}), there is no CPV cut
in the vertex correction to $N\to\ell H$ at zero temperature.
We do not discuss a generalization of the RIS subtraction
procedure to include finite temperature corrections (i.e.
the quantum statistical distributions of
the equilibrium particles $\ell$, $\widetilde \ell$, $H$, $\widetilde H$ and $\widetilde B$ in the present context), as we
find the derivation of the CPV rates based on the CTP approach
perhaps less heuristic and after all more intuitive and
technically simple.}
Such RIS with on-shell $H_1$ must be subtracted, and the rates
can be written in the suggestive way 
\begin{align}
\label{RIS:Idec}
\gamma^{\rm RIS}_{\ell\widetilde H\widetilde B\to N}
=&\gamma_{\widetilde H\widetilde B\to H_1}
\frac{Y_\ell}{Y_\ell^{\rm eq}}
\frac{\gamma^{\rm av}}{\gamma_{\widetilde H\widetilde B\to H_1}}
\frac{1-\varepsilon}{2}\,,\quad
\gamma^{\rm RIS}_{\ell\widetilde H\widetilde B\to N}
=&\gamma_{\widetilde H\widetilde B\to H_1}
\frac{Y_{\bar\ell}}{Y_\ell^{\rm eq}}
\frac{\gamma^{\rm av}}{\gamma_{\widetilde H\widetilde B\to H_1}}
\frac{1+\varepsilon}{2}\,,
\end{align}
where $\gamma_{\widetilde H\widetilde B\to H_1}$ is the rate
for the corresponding $2\leftrightarrow 1$ processes.
As we
assume supergauge interactions to be in equilibrium,
it is equal to the inverse rate, $\gamma_{\bar{\widetilde H}\widetilde B\to H_1^*}=\gamma_{\widetilde H\widetilde B\to H_1}$.

Note that even though the $1\leftrightarrow 3$ cut in Fig.~\ref{fig:offhiggswv} corresponds to the interference between $H_1$- and $H_2$-mediated exchange amplitudes with only $H_1$ being on-shell, it is nevertheless convenient to express it in terms of the rate for $H_1$ production, $\gamma_{\widetilde H\widetilde B\to H_1}$.
To see this, we cut the $H_1$-mediated
$1\leftrightarrow 3$ amplitude along the on-shell $H_1$ and fuse the
$\widetilde B$ and $\widetilde H$ lines with the $H_2$-mediated
$1\leftrightarrow 3$ amplitude. This way, we obtain an interference 
between a tree-level and a one-loop amplitude for $\widetilde H\widetilde B\to H_1$, cf. Fig.~\ref{fig:offhiggswv}. Consequently,
in the one-loop integral, only the cut contribution where
$\widetilde B$ and $\widetilde H$ are on-shell is extracted, which is
precisely the part that is relevant for CPV.
Note that this logic is similar to that of RIS subtraction
in conventional  leptogenesis, see e.g. Ref.~\cite{Beneke:2010wd}. In that work,  the RIS subtraction entails an interference between scattering amplitudes mediated by two different neutrinos $N_1$ and $N_2$ where only $N_1$ may be on-shell. The full rate is nevertheless characterized in terms of the rate for production of the $N_1$ (inverse) decay,  as can again be seen from cutting and fusing the scattering
amplitudes.

The second,
third and fourth factors on the right hand side combine
to the rate for an on-shell $H_1$ ($H_1^*$) to inversely
decay with $\ell$ ($\bar \ell$) into $N$, where we assume that
$\gamma_{\widetilde H\widetilde B\to H_1}\gg\gamma^{\rm av}$, such that $\gamma^{\rm av}/\gamma_{\widetilde H\widetilde B\to H_1}$ approximately is the branching ratio for $H_1$
inversely decaying with $\ell$ to $N$. Moreover,
\begin{align}
\label{RIS:dec}
\gamma^{\rm RIS}_{N\to\ell\widetilde H\widetilde B}
=\gamma^{\rm av}\frac{Y_N}{Y_N^{\rm eq}}\frac{1+\varepsilon}{2}
\,,\quad
\gamma^{\rm RIS}_{N\to\bar\ell\bar{\widetilde H}\widetilde B}
=\gamma^{\rm av}\frac{Y_N}{Y_N^{\rm eq}}\frac{1-\varepsilon}{2}
\,.
\end{align}
Substituting the rates~(\ref{RIS:Idec}) and~(\ref{RIS:dec}) into
Eq.~(\ref{BE:ql:softlepto}), the CPV
contributions cancel. The washout of a potentially
pre-existing lepton asymmetry is contained within the
unsubtracted, $CP$-conserving $1\leftrightarrow 3$ rates
[the 5th through 8th terms on the right hand side of
Eq.~(\ref{BE:ql:softlepto})]. One should notice that none of
the rates is explicitly weighted by factors of
$Y_{H_1,\widetilde H,\widetilde B}/Y_{H_1,\widetilde H,\widetilde B}^{\rm eq}$,
as we assume that $\widetilde H$ and $\widetilde B$ are in equilibrium.
Recall that this assumption is also crucial in order to establish
the results of vanishing asymmetries in
Sections~\ref{section:ltildeoff} and~\ref{section:wv}.

The main conclusion that may be drawn based on the present
discussion
of RIS subtraction and of the results in the other Sections
of this paper is that a complete network of kinetic equations
must properly account for all stable asymptotic states.
In particular, for standard leptogenesis,
diagrams with $N$ as an external
state have to be supplemented by corresponding diagrams
where $\ell$ and $H$ attach to $N$, i.e. the $1\leftrightarrow2$ (inverse) decay
processes have to be supplemented by $2\leftrightarrow2$ scatterings.
On-shell $\ell$ and $H$
yield also the crucial CPV contributions from loop
diagrams. Correspondingly, for the processes in
Sections~\ref{section:ltildeoff} and~\ref{section:wv}, the
kinetic equations must be augmented by attaching $\widetilde B$
and $\widetilde H$ to the external $H_1$ -- the same particles
that also give rise to CPV when they propagate on-shell
within a loop. In the latter case, the asymmetry vanishes, because
$H_1$ (the particle that attaches to the CPV loop involving $\widetilde B$
and $\widetilde H$) is in thermal equilibrium (due to gauge interactions), while for standard leptogenesis,
an asymmetry persists after the subtraction of RIS,
provided $N$ (the singlet neutrino
that attaches to the CPV loop involving $\ell$
and $H$)  is
out-of-equilibrium.

These conclusions can easily be generalized
to additional models. For example, the work of
Ref.~\cite{Hall:2010jx} included no RIS subtraction or any
other pertinent completion of the Boltzmann equations by attaching
loop particles to the decaying particle. As
the loop particles as well as the decaying particle are in thermal
equilibrium (all of them are gauged),
the resulting asymmetry from a complete set of kinetic 
equations vanishes.

\section{Conclusions}
\label{sec:conclusions}

In the study of novel, low-scale leptogenesis scenarios that may have interesting phenomenological consequences, it is essential to properly account for the unitary evolution of all the relevant states involved in the possible generation of a net lepton number asymmetry. In principle, one may do so following the conventional Boltzmann equation approach if one properly implements the RIS subtraction procedure and includes the statistical factors for all on-shell particles, whether they appear as explicit external states or as internal lines in loop graphs. The subtleties that this procedure entails makes RIS subtraction fraught with opportunities for error. Moreover, this approach appears somewhat heuristic, and a derivation of a proper inclusion of the
statistical factors of the internal line has not yet been
reported in the literature.
Alternatively, the CTP formulation provides a systematic approach that ensures unitary evolution and avoids the pitfalls one may encounter with the RIS subtraction. 

Using soft leptogenesis in $\nu$MSSM, where CPV is sourced through the phase $\mathrm{Arg}(\mu M_1 b^\ast)$, we have shown how the CTP approach with appropriate inclusion of all cuts includes all of the aforementioned requirements. With interactions analogous to those used  in Ref.~\cite{Fong:2009iu}, we then obtain a vanishing lepton number asymmetry, in contrast to the conclusion one would reach following the procedures taken in those calculations. Consequently, we argue that the conclusions reached in Ref.~\cite{Fong:2009iu} are unlikely to apply without extra input. Similar conclusions should apply to the asymmetric freeze-in scenario considered in Ref.~\cite{Hall:2010jx}\footnote{Concretely, in the diagrammatic example of Ref.~\cite{Hall:2010jx}, incoming gauge bosons and charginos or neutralinos should be attached to the decaying chargino. From the resulting scattering diagrams, the RIS should then be subtracted, such that the final asymmetry vanishes. To show this, one can follow the calculations presented Sections~\ref{section:ltildeoff} and~\ref{section:wv} of this present paper. Note that these calculations apply to all kinematic situations, provided the CPV cut is kinematically viable.}. The possibilities for low-scale, non-resonant leptogenesis nevertheless remain intriguing.  Exploration of the additional physics needed to make such scenarios viable will be the subject of forthcoming work.

\subsection*{Acknowledgments}
The authors are grateful to Apostolos Pilaftsis and
Daniel Chung for discussions regarding the importance
of RIS subtraction. BG acknowledges
support by the Gottfried Wilhelm Leibniz programme
of the Deutsche Forschungsgemeinschaft and by the DFG cluster of excellence `Origin and Structure of the Universe'.
MJRM was supported in part by U.S. Department of Energy
contract DE-FG02-08ER41531 and by the Wisconsin Alumni Research Foundation. MJRM also thanks the Excellence Cluster Universe and Technical University M\"unchen for their hospitality during completion of a portion of this work. 

\begin{appendix}

\section{Tree Level Propagators}
\label{appendix:propagators}

The results that are presented in this paper are obtained using
the CTP techniques for calculating CPV rates developed
in Ref.~\cite{Beneke:2010wd}. Additional papers aiming for the formulation
of kinetic theory based on the CTP approach
include~\cite{Calzetta:1986cq,Prokopec:2003pj,Prokopec:2004ic,Cirigliano:2009yt,Herranen:2010mh,Drewes:2010pf,Cirigliano:2011di,Herranen:2011zg,Fidler:2011yq,Garbrecht:2011xw,Tulin:2012re}.
Basic building blocks are the tree-level propagators. For scalar
particles, they take the form
\begin{subequations}
\label{prop:bose:expl}
\begin{align}
{\rm i}\Delta^<(p)&=
2\pi \delta(p^2-m^2)\left[
\vartheta(p_0) f(\mathbf p)
+\vartheta(-p_0) (1+\bar f(-\mathbf p))\right]
\,,
\\
{\rm i}\Delta^>(p)&=
2\pi \delta(p^2-m^2)\left[
\vartheta(p_0) (1+f(\mathbf p))
+\vartheta(-p_0) \bar f(-\mathbf p)\right]
\,,
\\
{\rm i}\Delta^T(p)&=
\frac{\rm i}{p^2-m^2+{\rm i}\varepsilon}+
2\pi \delta(p^2-m^2)\left[
\vartheta(p_0) f(\mathbf p)
+\vartheta(-p_0) \bar f(-\mathbf p)\right]
\,,
\\
{\rm i}\Delta^{\bar T}(p)&=
-\frac{\rm i}{p^2-m^2-{\rm i}\varepsilon}+
2\pi \delta(p^2-m^2)\left[
\vartheta(p_0) f(\mathbf p)
+\vartheta(-p_0) \bar f(-\mathbf p)\right]
\,.
\end{align}
\end{subequations}
In order to distinguish different fields, in the
present context $\widetilde \ell$ and $H_1$ nd $H_2$,
the Green functions $\Delta$, the masses $m$ and the
particle and antiparticle
distribution functions $f$ and $\bar f$ are marked with subscripts.
${\rm SU}(2)$ gauge group indices are suppressed.

For spin-$1/2$ fermions, the Green functions are
\begin{subequations}
\label{prop:fermi:expl}
\begin{align}
{\rm i}S^{<}(p)
&=-2\pi\delta(p^2-m^2)(p\!\!\!/+m)\left[
\vartheta(p_0)f(\mathbf{p})
-\vartheta(-p_0)(1-\bar f(-\mathbf{p}))
\right]\,,\\
{\rm i}S_{Ni}^{>}(p)
&=-2\pi\delta(p^2-m^2)(p\!\!\!/+m)\left[
-\vartheta(p_0)(1-f(\mathbf{p}))
+\vartheta(-p_0)\bar f(-\mathbf{p})
\right]\,,\\
{\rm i}S^{T}(p)
&=
\frac{{\rm i}(p\!\!\!/+M_i)}{p^2-M_i^2+{\rm i}\varepsilon}
-2\pi\delta(p^2-m^2)(p\!\!\!/+m)\left[
\vartheta(p_0)f(\mathbf{p})
+\vartheta(-p_0)\bar f(-\mathbf{p})
\right]\,,\\
{\rm i}S^{\bar T}(p)
&=
-\frac{{\rm i}(p\!\!\!/+m)}{p^2-m^2-{\rm i}\varepsilon}
-2\pi\delta(p^2-M_i^2)(p\!\!\!/+m)\left[
\vartheta(p_0)(\mathbf{p})
+\vartheta(-p_0)\bar f(-\mathbf{p})
\right]
\,.
\end{align}
\end{subequations}
Again, subscripts distinguish the fields $\ell$, $\widetilde H$
and $\widetilde B$ and ${\rm SU}(2)$ indices are suppressed.
Majorana fermions observe the constraint
$f(\mathbf p)=\bar f(\mathbf p)$.

\section{Replacement of Time-Ordered by On-Shell Green Functions}
\label{appendix:replacements}

We consider the integral
\begin{align}
P=&\int\frac{d^4 k}{(2\pi)^4}
\left[
{\rm i}\Delta^T_{X1}(p+k)
{\rm i}\Delta^T_{X2}(q+k)
+
{\rm i}\Delta^{\bar T}_{X1}(p+k)
{\rm i}\Delta^{\bar T}_{X2}(q+k)
\right]
g(k)
\\\notag
=&
\int\frac{d^4 k}{(2\pi)^4}
\Bigg[
\frac{\rm i}{(p+k)^2-m_1^2+{\rm i}\varepsilon}
\\\notag
&
+2\pi \delta\left((p+k)^2-m_1^2\right)\left[
\vartheta(p_0+k_0) f_{X1}(\mathbf p+\mathbf k)
+\vartheta(-p_0-k_0) \bar f_{X1}(-\mathbf p-\mathbf k)\right]
\Bigg]
\\\notag
&\times\Bigg[
\frac{\rm i}{(q+k)^2-m_2^2+{\rm i}\varepsilon}
\\\notag
&
+2\pi \delta\left((q+k)^2-m_2^2\right)\left[
\vartheta(q_0+k_0) f_{X2}(\mathbf q+\mathbf k)
+\vartheta(-q_0-k_0) \bar f_{X2}(-\mathbf q-\mathbf k)\right]
\Bigg]
g(k)
\\\notag
+&\int\frac{d^4 k}{(2\pi)^4}
\Bigg[
-\frac{\rm i}{(p+k)^2-m_1^2-{\rm i}\varepsilon}
\\\notag
&
+2\pi \delta\left((p+k)^2-m_1^2\right)\left[
\vartheta(p_0+k_0) f_{X1}(\mathbf p+\mathbf k)
+\vartheta(-p_0-k_0) \bar f_{X1}(-\mathbf p-\mathbf k)\right]
\Bigg]
\\\notag
&\times\Bigg[
-\frac{\rm i}{(q+k)^2-m_2^2-{\rm i}\varepsilon}
\\\notag
&
+2\pi \delta\left((q+k)^2-m_2^2\right)\left[
\vartheta(q_0+k_0) f_{X2}(\mathbf q+\mathbf k)
+\vartheta(-q_0-k_0) \bar f_{X2}(-\mathbf q-\mathbf k)\right]
\Bigg]
g(k)\,.
\end{align}
Integrals of this type are encountered throughout the calculations
of the collision terms ${\cal C}$
in the present work. In first place, the individual terms include only
products of either two time-ordered or two anti-time ordered propagators,
not their sum. However, the terms convoluted with
the product of two $\bar T$-propagators in the integrated
two-loop collision terms
can generally be brought to the form of the terms convoluted with
the two $T$-propagators by reversing the sign of all momentum variables,
making use of ${\rm i}\Delta^{<,>}_{X}(k)={\rm i}\Delta^{>,<}_{X}(-k)$
and ${\rm i}\Delta^{T,\bar T}_{X}(k)={\rm i}\Delta^{T,\bar T}_{X}(-k)$, for
isotropic, charge neutral distributions.
In this Appendix, we explain the
replacement rule for combinations of scalar fields, but corresponding results
are easily seen to follow for integrals involving fermionic fields as well (in which
case one has also to pay attention to the behavior of the spinor structure
under sign reversals).

The terms containing products of finite-density
contributions combine to
\begin{align}
P_1=&2\int\frac{d^4 k}{(2\pi)^4}
\Bigg[2\pi \delta\left((p+k)^2-m_1^2\right)
\left[
\vartheta(p_0+k_0) f_{X1}(\mathbf p+\mathbf k)
+\vartheta(-p_0-k_0) \bar f_{X1}(-\mathbf p-\mathbf k)\right]
\Bigg]
\\\notag
\times&
\Bigg[
2\pi \delta\left((q+k)^2-m_2^2\right)
\left[
\vartheta(q_0+k_0) f_{X2}(\mathbf q+\mathbf k)
+\vartheta(-q_0-k_0) \bar f_{X2}(-\mathbf q-\mathbf k)
\right]
\Bigg]
g(k)\,.
\end{align}
For the products of finite density and vacuum terms one obtains
\begin{align}
P_2=&\int\frac{d^4 k}{(2\pi)^4}
2\pi \delta\left((p+k)^2-m_1^2\right)
\left[
\vartheta(p_0+k_0) f_{X1}(\mathbf p+\mathbf k)
+\vartheta(-p_0-k_0) \bar f_{X1}(-\mathbf p-\mathbf k)
\right]
\\\notag
\times&2\pi \delta\left((q+k)^2-m_1^2\right)g(k)
\\\notag
+&\int\frac{d^4 k}{(2\pi)^4}
2\pi \delta\left((q+k)^2-m_2^2\right)
\left[
\vartheta(q_0+k_0) f_{X2}(\mathbf p+\mathbf k)
+\vartheta(-q_0-k_0) \bar f_{X2}(-\mathbf p-\mathbf k)
\right]
\\\notag
\times&2\pi \delta\left((q+k)^2-m_2^2\right)g(k)
\,.
\end{align}
Finally, the products of the vacuum contributions yield
\begin{align}
P_3=&\int\frac{d^4 k}{(2\pi)^4}
\left[
\frac{\rm i}{(p+k)^2-m_1^2+{\rm i}\varepsilon}
\frac{\rm i}{(q+k)^2-m_2^2+{\rm i}\varepsilon}
+\frac{(-{\rm i})}{(p+k)^2-m_1^2-{\rm i}\varepsilon}
\frac{(-{\rm i})}{(q+k)^2-m_2^2-{\rm i}\varepsilon}
\right]
g(k)
\\\notag
=&\int\frac{d^4 k}{(2\pi)^4}
2\pi \delta\left((p+k)^2-m_1^2\right)
2\pi \delta\left((q+k)^2-m_2^2\right)
g(k)\,.
\end{align}
Note that this
contribution gives, of course, the same result as obtained by applying the
vacuum cutting rules.

In summary, we obtain
\begin{align}
P=&P_1+P_2+P_3
=\int\frac{d^4 k}{(2\pi)^4}
2\pi \delta\left((p+k)^2-m_1^2\right)
2\pi \delta\left((q+k)^2-m_2^2\right)
\\\notag
\times&
\Bigg\{
1+\left[
\vartheta(p_0+k_0) f_{X1}(\mathbf p+\mathbf k)
+\vartheta(-p_0-k_0) \bar f_{X1}(-\mathbf p-\mathbf k)
\right]
\\\notag
&+\left[
\vartheta(q_0+k_0) f_{X2}(\mathbf p+\mathbf k)
+\vartheta(-q_0-k_0) \bar f_{X2}(-\mathbf p-\mathbf k)
\right]
\\\notag
&+2
\left[
\vartheta(p_0+k_0) f_{X1}(\mathbf p+\mathbf k)
+\vartheta(-p_0-k_0) \bar f_{X1}(-\mathbf p-\mathbf k)
\right]
\\\notag
&\times
\left[
\vartheta(q_0+k_0) f_{X2}(\mathbf p+\mathbf k)
+\vartheta(-q_0-k_0) \bar f_{X2}(-\mathbf p-\mathbf k)
\right]
\Bigg\}
g(k)\,.
\end{align}

Now, provided at the point where the on-shell $\delta$-functions
are simultaneously fulfilled, ${\rm sign}(p_0+k_0)={\rm sign}(q_0+k_0)$,
we may express
\begin{align}
P=&P_1+P_2+P_3
=\int\frac{d^4 k}{(2\pi)^4}
\left[
{\rm i}\Delta_{X1}^>(p+k){\rm i}\Delta_{X2}^>(q+k)
+{\rm i}\Delta_{X1}^<(p+k){\rm i}\Delta_{X2}^<(q+k)
\right]
g(k)\,.
\end{align}
In the other case, ${\rm sign}(p_0+k_0)=-{\rm sign}(q_0+k_0)$,
it is
\begin{align}
P=&P_1+P_2+P_3
=\int\frac{d^4 k}{(2\pi)^4}
\left[
{\rm i}\Delta_{X1}^>(p+k){\rm i}\Delta_{X2}^<(q+k)
+{\rm i}\Delta_{X1}^<(p+k){\rm i}\Delta_{X2}^>(q+k)
\right]
g(k)\,.
\end{align}

\section{Evaluation of the Collision Terms in Different Kinematic Situations}
\label{appendix:coll:kinematics}

On the example of the contributions to the collision term
from off-shell binos, that is computed in Section~\ref{sec:offshellbino}, we show how
to further evaluate it in different kinematic situations, i.e.
when $m_N>m_{\widetilde \ell}+m_{H1}$ and $m_N+m_{H1}<m_{\widetilde \ell}$. (We assume that $H_1$ is lighter than $N$ and $\widetilde \ell$.)
While this is not necessary for the main purpose here, which is to show the cancellation
${\cal C}^{{\rm v}\widetilde B}_{\widetilde \ell}(\mathbf k)+{\cal C}^{{\rm v}\widetilde B}_{\ell}(\mathbf k)=0$, it is instructive to show how the collision terms
are related to the usual CPV source terms in the Boltzmann equations
that are proportional to $\delta f_N$. We therefore concentrate
on ${\cal C}^{{\rm v}\widetilde B}_{\widetilde \ell}(\mathbf k)$
and
further simplify this expression by performing an integration over
$d^3k$. Then, we need to distinguish the cases
$m_N>m_{\widetilde \ell}+m_{H1}$ and $m_N+m_{H1}<m_{\widetilde \ell}$.\footnote{Note that in the latter case, there would be no cut contribution in the vacuum,
but there is one present at finite temperature. In the context of standard Leptogenesis
this is pointed out and calculated in Refs.~\cite{Giudice:2003jh,Garbrecht:2010sz}.}
This is because the
$\vartheta$-functions occurring within the expressions for the finite-density propagators
effectively distinguish between these situations. First, for $m_N>m_{\widetilde \ell}+m_{H1}$,
we find
\begin{align}
\label{source:ltilde:heavyN}
\int \frac{d^3k^\prime}{(2\pi)^3}{\cal C}^{{\rm v}\widetilde B}_{\widetilde \ell}({\mathbf k}^\prime)
&=-Y^2g_1^2\sin\phi_\mu\sin\alpha\cos\alpha
\int\frac{d^3 k}{(2\pi)^32\sqrt{{\mathbf k}^2+m_N^2}}
\delta f_N(\mathbf k) k^\mu
\\\notag
&
\hskip-2.0cm\times
\int\frac{d^3 k^\prime}{(2\pi)^3 2\sqrt{{{\mathbf k}^\prime}^2+m_{\widetilde \ell}^2}}
\frac{d^3 k^{\prime\prime}}{(2\pi)^3 2\sqrt{{{\mathbf k}^{\prime\prime}}^2+\mu^2}}
(2\pi)^4\delta^4(k-k^\prime-k^{\prime\prime})
\left[1-f_{\widetilde H}({\mathbf{k}}^{\prime\prime})
+f_{\widetilde \ell}(\mathbf{k}^\prime)\right]
\\\notag
&\hskip-2.0cm\times
\int\frac{d^3 p}{(2\pi)^3 2\sqrt{{{\mathbf p}}^2}}
\frac{d^3 p^{\prime}}{(2\pi)^3 2\sqrt{{{\mathbf p}^{\prime}}^2+m_{H1}^2}}
p_\mu
(2\pi)^4\delta^4(k-p^\prime-p)
\left[1+f_{H1}({\mathbf{p}}^{\prime})
-f_{\ell}(\mathbf{p})\right]
\\\notag
&\hskip7.5cm
\times
\frac{\mu M_1}{(p+k^\prime)^2-M_1^2}
\,.
\end{align}
When $m_N+m_{H1}<m_{\widetilde \ell}$, the result is
\begin{align}
\label{source:ltilde:lightN}
\int \frac{d^3k^\prime}{(2\pi)^3}{\cal C}^{{\rm v}\widetilde B}_{\widetilde \ell}({\mathbf k}^\prime)
&=-Y^2g_1^2\sin\phi_\mu\sin\alpha\cos\alpha
\int\frac{d^3 k}{(2\pi)^32\sqrt{{\mathbf k}^2+m_N^2}}
\delta f_N(\mathbf k) k^\mu
\\\notag
&
\hskip-2.0cm\times
\int\frac{d^3 k^\prime}{(2\pi)^3 2\sqrt{{{\mathbf k}^\prime}^2+m_{\widetilde \ell}^2}}
\frac{d^3 k^{\prime\prime}}{(2\pi)^3 2\sqrt{{{\mathbf k}^{\prime\prime}}^2+\mu^2}}
(2\pi)^4\delta^4(k-k^\prime+k^{\prime\prime})
\left[f_{\widetilde H}({\mathbf{k}}^{\prime\prime})
+f_{\widetilde \ell}(\mathbf{k}^\prime)\right]
\\\notag
&\hskip-2.0cm\times
\int\frac{d^3 p}{(2\pi)^3 2\sqrt{{{\mathbf p}}^2}}
\frac{d^3 p^{\prime}}{(2\pi)^3 2\sqrt{{{\mathbf p}^{\prime}}^2+m_{H1}^2}}
p_\mu
(2\pi)^4\delta^4(k-p^\prime-p)
\left[1+f_{H1}({\mathbf{p}}^{\prime})
-f_{\ell}(\mathbf{p})\right]
\\\notag
&\hskip7.5cm\times
\frac{\mu M_1}{(p+k^\prime)^2-M_1^2}
\,.
\end{align}
The difference between Eq.~(\ref{source:ltilde:heavyN}) and
Eq.~(\ref{source:ltilde:lightN}) is within the statistical weights
of $\widetilde H$ and $\widetilde \ell$. As one should anticipate,
Eq.~(\ref{source:ltilde:lightN}) vanishes in the vacuum, where all distribution functions
are {\it zero}. In the reminder of this
work, we do not distinguish between the different 
kinematic possibilities, because this is not necessary in order to readily
see the cancellations that are of relevance for soft leptogenesis.

\end{appendix}

\end{document}